%% file: main.tex
\documentclass[sigconf,screen,authorversion]{acmart}
\PassOptionsToPackage{usenames,dvipsnames}{xcolor}

\usepackage{balance}
\usepackage{microtype}
\usepackage{etex}
\usepackage{amsmath}
\usepackage{amsfonts}
\usepackage{algorithmic}
\usepackage{graphicx}
\usepackage{textcomp}
\usepackage{fancyhdr}
\usepackage{tabu}
\usepackage{enumitem}
\usepackage[normalem]{ulem}
\usepackage[scaled]{beramono}
\usepackage{arydshln}
\usepackage{caption}
\usepackage{subcaption}
\usepackage{adjustbox}
\usepackage{etoolbox}
\usepackage{verbatim}   
\usepackage{wrapfig}
\usepackage{multirow}
\usepackage{balance}
\usepackage{pifont}

\usepackage{rotating}

\usepackage{xspace} 
\newcommand{\SCHEDULER}{Aquatope\xspace}

\begin{document}

\title{QoS-Aware Resource Management for Multi-phase Serverless Workflows with Aquatope}

\author{Zhuangzhuang Zhou}
\affiliation{%
  \institution{Cornell University}
  \city{Ithaca}
  \state{New York}
  \country{USA}
  \postcode{14850}
}
\email{zz586@cornell.edu}

\author{Yanqi Zhang}
\affiliation{%
  \institution{Cornell University}
  \city{Ithaca}
  \state{New York}
  \country{USA}
  \postcode{14850}
}
\email{yz2297@cornell.edu}

\author{Christina Delimitrou}
\affiliation{%
  \institution{MIT}
  \city{Cambridge}
  \state{Massachusetts}
  \country{USA}
  \postcode{02139}
}
\email{delimitrou@csail.mit.edu}

\date{}

\begin{abstract}
\input{Abstract.tex}
\end{abstract}




\maketitle

\input{defs}

\input{Introduction.tex}

\input{Background.tex}

\input{Design.tex}

\input{Implementation.tex}

\input{Methodology.tex}

\input{Evaluation.tex}


\input{Conclusions.tex}


\balance

\bibliographystyle{ACM-Reference-Format}
\bibliography{references}

\end{document}

%% file: Abstract.tex
Multi-stage serverless applications, i.e., workflows with many computation and I/O stages, are becoming increasingly representative of FaaS platforms. Despite their advantages in terms of fine-grained scalability and modular development, these applications are subject to suboptimal performance, resource inefficiency, and high costs to a larger degree than previous simple serverless functions.

We present Aquatope, a QoS-and-uncertainty-aware resource scheduler for end-to-end serverless workflows that takes into account the inherent uncertainty present in FaaS platforms, and improves performance predictability and resource efficiency. Aquatope uses a set of scalable and validated Bayesian models to create pre-warmed containers ahead of function invocations, and to allocate appropriate resources at function granularity to meet a complex workflow's end-to-end QoS, while minimizing resource cost. Across a diverse set of analytics and interactive multi-stage serverless workloads, Aquatope significantly outperforms prior systems, reducing QoS violations by $5\times$, and cost by 34\% on average and up to 52\% compared to other QoS-meeting methods.

%% file: defs.tex
\newtoggle{showmarks}

\iftoggle{showmarks}{
  \newcommand\yanqi[1]{\textcolor{blue}{YANQI: #1}}
  \newcommand\zz[1]{\textcolor{blue}{ZZ: #1}}
  \newcommand\christina[1]{\textcolor{red}{CHRISTINA: #1}}
}{
  \newcommand\zz[1]{\unskip}
  \newcommand\yanqi[1]{\unskip}
  \newcommand\christina[1]{\unskip}
}

%% file: Introduction.tex
\section{Introduction}
\label{sec:introduction}

Serverless computing is becoming increasingly popular, due to its ease of programming and maintenance, fast elasticity, and fine-grained billing. Serverless simplifies management for users, since its interface removes the need for users to explicitly configure virtual machines (VMs) or containers. Serverless also avoids overprovisioning, as users only pay for the resources they use during execution. For applications with high data-level parallelism and intermittent activity, serverless can achieve much higher performance for the same or lower cost.

Despite these benefits, serverless introduces several challenges, especially when a service has to meet quality of service (QoS) requirements in terms of execution time or tail latency~\cite{Delimitrou13,Delimitrou13c,Delimitrou13d,Delimitrou14,Delimitrou15,Delimitrou16,Delimitrou19,gan2019asplos,Lazarev21}. A lot of prior work has focused on reducing the cold start overheads in serverless, i.e., overheads associated with instantiating new containers or VMs, and installing necessary dependencies~\cite{wang18peeking, du20catalyzer,shahrad2020serverless,ustiugov2021snapshots, fuerst2021faascache}. While impactful, cold starts are not the sole reason behind degraded performance in serverless. Another crucial issue the system has to tackle is appropriate function-level resource management. Without a proper resource configuration, the function can suffer from performance degradation and increased execution cost~\cite{yu2020serverless}. More importantly, these two problems are closely correlated with each other, as cold and warm starts lead to different function performance, and require significantly different resources. For the system to minimize cost, while satisfying QoS, we need to tackle both challenges jointly.

Furthermore, serverless providers are increasingly providing workflow programming model interfaces, where each serverless application consists of multiple loosely-coupled functions in pursuit of fine-grained scalability and modular development and deployment~\cite{awsstep,azuredurable,yu2020serverless}. 
The challenges above are amplified for multi-stage serverless workflows, where cascading cold starts across dependent stages~\cite{nilanjan20xanadu} and varied resource needs for each stage~\cite{yu2020serverless} make cold start elimination and resource management even more challenging.
Finally, serverless is prone to high system-level noise due to the interference from colocated workloads in FaaS deployments, which further hinder performance predictability~\cite{shahrad2020serverless}. 

We present \SCHEDULER, a QoS-and-uncertainty-aware scheduler for multi-stage serverless workloads that jointly tackles the two main challenges contributing to degraded performance and inefficiency in Function-as-a-Service (FaaS): cold starts and function-level resource allocation. \SCHEDULER consists of two major components, \textit{a dynamic pre-warmed container pool} and \textit{a container resource manager}. The dynamic pre-warmed container pool uses a hybrid Bayesian neural network to adjust the number of pre-warmed containers. The container resource manager leverages Bayesian Optimization to search for a near-optimal resource configuration for each execution stage in a workflow. \SCHEDULER uses a Bayesian approach to account for the noise and uncertainty that are prevalent in FaaS platforms due to stochasticity inherent to function execution, load fluctuation, and interference from colocated applications. \SCHEDULER is a centralized controller, operates online, transparently to the user, and introduces marginal overheads.

We implement \SCHEDULER on OpenWhisk~\cite{openwhisk} and evaluate it across a wide set of analytics and interactive multi-stage serverless applications, including ML pipelines, video processing frameworks, and social networks. In all cases, \SCHEDULER outperforms prior empirical and ML-driven approaches in performance and efficiency, reducing QoS violations by $5\times$ compared to prior work, and execution cost by 34\% on average and up to 52\% compared to other QoS-meeting methods.

%% file: Background.tex
\section{Background and Motivation}
\label{sec:background}

\subsection{Problem Statement}
\label{sec:problem_statement}
 
Many real-world serverless applications are implemented as \textit{multi-stage serverless workflow} in which incoming user requests invoke sets of serverless functions that coordinate with each other to execute a serverless workflow. Existing platforms provide various composition mechanisms to control a workflow and transfer intermediate state across functions~\cite{composer, awsstep, azuredurable}. By splitting a complex application into dependent but loosely-coupled functions, the application benefits from fine-grained scalability, parallel execution, and modular development~\cite{yu2020serverless}.
At the same time, decoupling a serverless application into multiple stages also introduces challenges in resource management, including additional function instantiation overheads, data transfer overheads, and varied resource needs across execution stages. Without an appropriate framework in place, multi-stage serverless workflows can experience QoS violations and resource inefficiency. 
 
\SCHEDULER specifically targets such serverless workflows that must meet pre-defined QoS constraints. \SCHEDULER tackles two correlated aspects of serverless resource management: ensuring that function instantiation overheads are minimal so that tasks do not suffer from cold starts, and optimizing the resource configuration of each stage to minimize cost while satisfying QoS. While \SCHEDULER is geared towards multi-stage serverless workloads, it can also be applied to simpler applications with a single stage.

\subsection{Challenges}
\label{sec:challeges}

Resource management for multi-stage serverless workflows faces the following challenges.

\vspace{0.08in}
\noindent{\bf{Cold starts:}}
Cold starts are one of the most studied overheads associated with serverless~\cite{wang18peeking,shahrad2019arch,yu2020serverless,fuerst2021faascache,roy2022icebreaker}. A \textit{cold start} invocation occurs when a serverless application is triggered, but its function instances are not yet loaded in memory. For the FaaS platform, a cold start involves launching a new container (and/or a new VM), setting up its runtime environment, and fetching and loading necessary libraries and dependencies. This process can take a long time relative to the short-lived function execution~\cite{wang18peeking,fuerst2021faascache}.

\vspace{0.08in}
\noindent{\bf{Diverse resource requirements:}} 
Serverless functions vary in functionality and are implemented with different libraries and runtimes. Their resource requirements also vary a lot~\cite{shahrad2019arch, yu2020serverless}, and without proper resource management, both performance and cost can suffer. Existing FaaS platforms, including AWS Lambda~\cite{awslambda}, Google Cloud Functions~\cite{googlefunction}, and IBM Cloud Functions~\cite{ibmfunction} require users to specify a memory limit for serverless functions, and allocate CPU resources proportional to the amount of provisioned memory, which can lead to CPU or memory overprovisioning.

\vspace{0.08in}
\noindent{\bf{Correlation of cold start and resource allocation:}} 
Cold starts not only affect the function startup latency but also exacerbate runtime performance degradation, as they can prevent a function invocation from reusing its execution context~\cite{awspractices}, which caches global variables (e.g., SDK clients, database connections, ML models, etc.). In this case, the function is forced to execute the user-provided initialization code to download data dependencies and initialize runtime packages, etc.~\cite{fuerst2021faascache}. This leads to different runtime performance and resource requirements for warm and cold starts, with cold start function invocations requiring more resources to meet the same performance target than warm start invocations. 
Without eliminating cold starts, the resource manager is forced to strike a balance between the performance behaviors of cold and warm starts, leading to degraded performance and excess resources. 

\vspace{0.08in}
\noindent{\bf{Multi-stage serverless workflow overheads:}}
Serverless application developers tend to decouple complex applications (e.g., ML inference, interactive web service) into workflows of loosely-coupled functions. Despite the advantages of fine-grained scalability and modular development, the performance of such applications can suffer for multiple reasons. First, the startup overhead is amplified by cascading cold starts across dependent functions~\cite{nilanjan20xanadu, yu2020serverless}.
Second, resource requirements can vary a lot across the execution stages of the same workflow.
Without proper resource management for each execution stage, the application would either fail to satisfy its QoS and/or suffer from increased cost. Additionally, different function composition methods (e.g., asynchronous invocation, function callback, function chaining, fan-in/fan-out, etc.) introduce more performance unpredictability, which makes finding a near-optimal resource configuration for the whole application more challenging.

\vspace{0.08in}
\noindent{\bf{Uncertainty in FaaS:}}
Noise and uncertainty are inherent to FaaS platforms. Serverless is well-suited for applications with fluctuating workloads due to their fine-grained scalability and pay-as-you-go pricing model~\cite{awslambda}. A large fraction of serverless applications have significant variability in invocation patterns, making it difficult to provision appropriate resources for them in advance~\cite{shahrad2020serverless, fuerst2021faascache}. In addition, due to their short execution time and fine granularity, cloud providers tend to colocate serverless functions to higher degrees than traditional cloud services. As a result, functions can suffer from interference from colocated workloads and lead to unpredictable performance~\cite{wang18peeking, yu2020serverless}, which causes biased observations and impairs the performance of sampling-based resource management approaches~\cite{patel2020clite, li2021rambo, roy2021satori}.

\subsection{Related Work}

\vspace{0.08in}
\noindent{\bf{Mitigating cold starts:}}
Cold start overheads have been studied extensively, including pre-crafting virtual network interfaces~\cite{mohan2019agile}, restoring a function from a well-formed checkpoint image to skip initialization~\cite{du20catalyzer}, and prefetching a function's working set of memory pages~\cite{ustiugov2021snapshots}. Most FaaS providers keep container instances loaded in memory for a fixed amount of time after a function terminates~\cite{wang18peeking}. AWS Lambda offers a \textit{provisioned concurrency}~\cite{provisioned} configuration to pre-load a fixed number of containers to accelerate function startup. FaaSCache~\cite{fuerst2021faascache} uses a caching-inspired container eviction policy to terminate containers when the server is saturated, but does not pre-warm containers. Shahrad et al.~\cite{shahrad2020serverless} proposed a histogram-based policy to adjust a container's keep-alive time. Similarly, IceBreaker~\cite{roy2022icebreaker} uses a Fourier-transformation-based model to predict future invocation patterns, and pre-warms function containers accordingly. These techniques can mitigate cold starts, but are often not robust to fluctuating workloads, and are designed for single-stage serverless applications, which are not prone to cascading cold starts across dependent functions. \SCHEDULER is complementary to these proposals but focuses on reducing cold starts for realistic, multi-stage serverless applications. 

\vspace{0.08in}
\noindent{\bf{Resource scheduling for FaaS:}}
Many FaaS resource schedulers focus on the storage side of serverless. Pocket~\cite{klimovic2018pocket} uses user-provided workload hints to rightsize storage resources. Pu et al.~\cite{pu2019shuffling} build application-specific performance models to select the storage configuration that achieves the desired cost-performance trade-off. There are a few systems that address the compute side of serverless management. Saha et al.~\cite{saha2018emars} and Suresh et al.~\cite{suresh2020ensure} use autoscaling to adjust a container's memory to satisfy a function's latency requirements. These systems are again designed for single-stage applications, and do not handle the diverse needs of different execution stages.

\vspace{0.08in}
\noindent{\bf{QoS-aware cloud management:}}
There has been extensive work on resource managers that meet QoS for latency-critical cloud applications~\cite{Delimitrou13,Delimitrou14,Mars13a,Zhang21,Gan21,Delimitrou19}. PARTIES~\cite{chen2019parties} showed that resources of interactive services are fungible, which simplifies resource partitioning when colocating multiple latency-critical jobs. CLITE~\cite{patel2020clite}, RAMBO~\cite{li2021rambo}, and SATORI~\cite{roy2021satori} showed that Bayesian Optimization (BO) can identify resource configurations that meet QoS for latency-critical jobs, maximize throughput for batch workloads, and preserve fairness among colocated jobs. While these systems improve performance and resource efficiency, they are designed for long-running applications, and cannot be directly applied to multi-stage serverless applications built with transient function containers. Moreover, these approaches do not consider the noise and uncertainty present in FaaS infrastructures, which can greatly hinder traditional BO techniques.

%% file: Design.tex
\section{\SCHEDULER Design Overview}
\label{sec:design}

\begin{figure}[t]
    \centering 
    \includegraphics[width=\columnwidth]{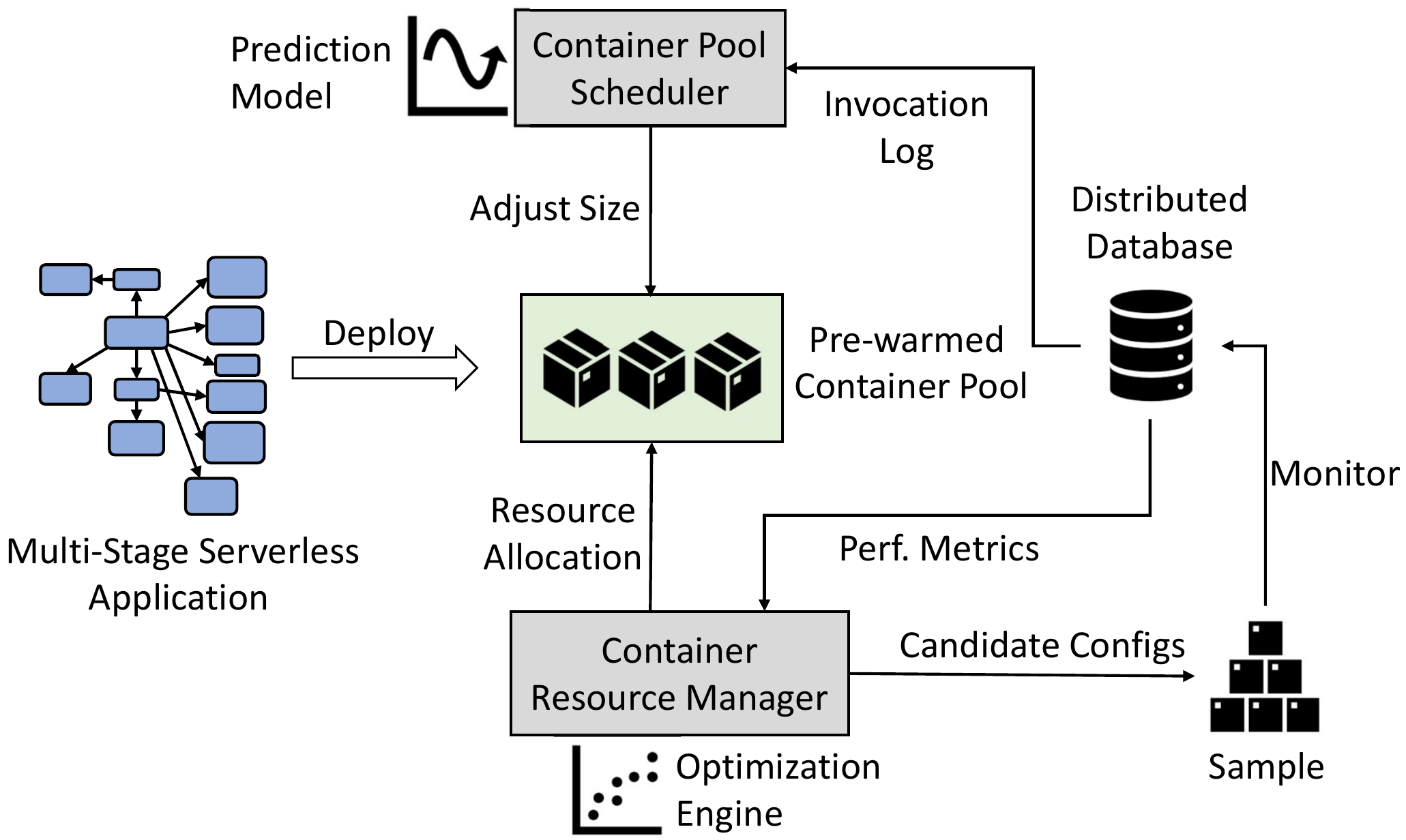}
	\caption{System overview of \SCHEDULER. }
    \label{fig:system_overview}
\end{figure}

\SCHEDULER is a QoS-and-uncertainty-aware resource scheduler for end-to-end, multi-stage serverless workflows. The design objective of \SCHEDULER is to meet the user-defined QoS of a multi-stage serverless application, while using the minimum amount of resources. To this end, \SCHEDULER jointly tackles two correlated challenges in serverless: (1) it maintains a \textit{dynamic pre-warmed container pool} to minimize cold starts and ensure most of the function invocations are handled by warm containers, and (2) it employs a \textit{container resource manager} that allocates appropriate resources to each function, based on its warm-start performance behavior.

Both components are implemented with Bayesian approaches to overcome the uncertainties present in FaaS platforms, including workload fluctuations, performance unpredictability, and interference from colocated applications. The dynamic pre-warmed container pool uses a hybrid Bayesian neural network~\cite{zhang2000neural}, which provides accurate and high-confidence predictions of function invocation rates, allowing \SCHEDULER to adjust the number of pre-warmed containers ahead of function invocations. The container resource manager then builds surrogate models to approximate the relationship between resource configurations and end-to-end performance and cost. The engine uses customized Bayesian optimization to efficiently explore the resource allocation space to find a configuration that meets the end-to-end QoS with minimal overprovisioning. It arrives at a suitable configuration by balancing exploration and exploitation, while adapting to noise in the cloud.

By integrating these two components, \SCHEDULER jointly tackles both challenges, and allows multi-stage serverless applications to operate in a performant and efficient manner. The following two sections describe each of \SCHEDULER's components in detail. 

\input{Design_LSTM.tex}

\input{Design_BO.tex}

%% file: Design_LSTM.tex
\section{Eliminating Cold Starts}

\SCHEDULER maintains a pool of pre-warmed containers to handle incoming function invocations. \SCHEDULER sizes the pre-warmed container pool at runtime such that there are just enough warm containers to handle incoming function invocations. \SCHEDULER also determines when to terminate a function's container to reclaim unused resources. 

Since multi-stage serverless applications are built with diverse runtimes and topologies, the optimal number of pre-warmed containers is application-specific. \SCHEDULER uses a set of machine learning (ML) models to infer the total number of required containers for each active serverless application over the next time interval, and adjusts the number of different types of containers accordingly. While several models can be applied towards this purpose, \SCHEDULER uses a hybrid Bayesian neural network to infer future invocation rates, which achieves high accuracy, fast inference, and agility to load fluctuations.

\subsection{Time Series Prediction} 

The problem of predicting function invocation patterns can be formalized as follows: given a number of different types of active function containers for a serverless workflow in the past $t$ time windows $\{{x_1}, {x_2}, ..., {x_t}\}$, we need to predict the invocation pattern for the next time window $\{x_{t+1}\}$. The time window size is configurable, and is set to 1 minute by default, which is the typical timescale for container keep-alive times in FaaS platforms~\cite{shahrad2020serverless, openwhisk}.
External features, which are the time of day, time of week, and function trigger types (HTTP, object storage, event hub, etc.), also need to be integrated into the prediction model to improve accuracy.
\SCHEDULER also accounts for the dependencies between functions in a multi-stage workflow, by predicting the invocation pattern of downstream containers in $x_{t+1}$, when it sees their upstream containers invoked in $\{{x_1}, {x_2}, ..., {x_t}\}$.
This captures both probabilistic and deterministic dependencies between execution stages, by predicting the expected and exact number of containers respectively. 
Since load fluctuates and invocation patterns may change, it is also important to incorporate uncertainty estimation to improve the robustness of the model, and to ensure that the scheduler makes reliable decisions and can recover from anomalies.

\subsection{Hybrid Bayesian Neural Network Model} 
\label{sec:hybrid_model}

\begin{figure}[t]
    \centering 
    \includegraphics[width=\columnwidth]{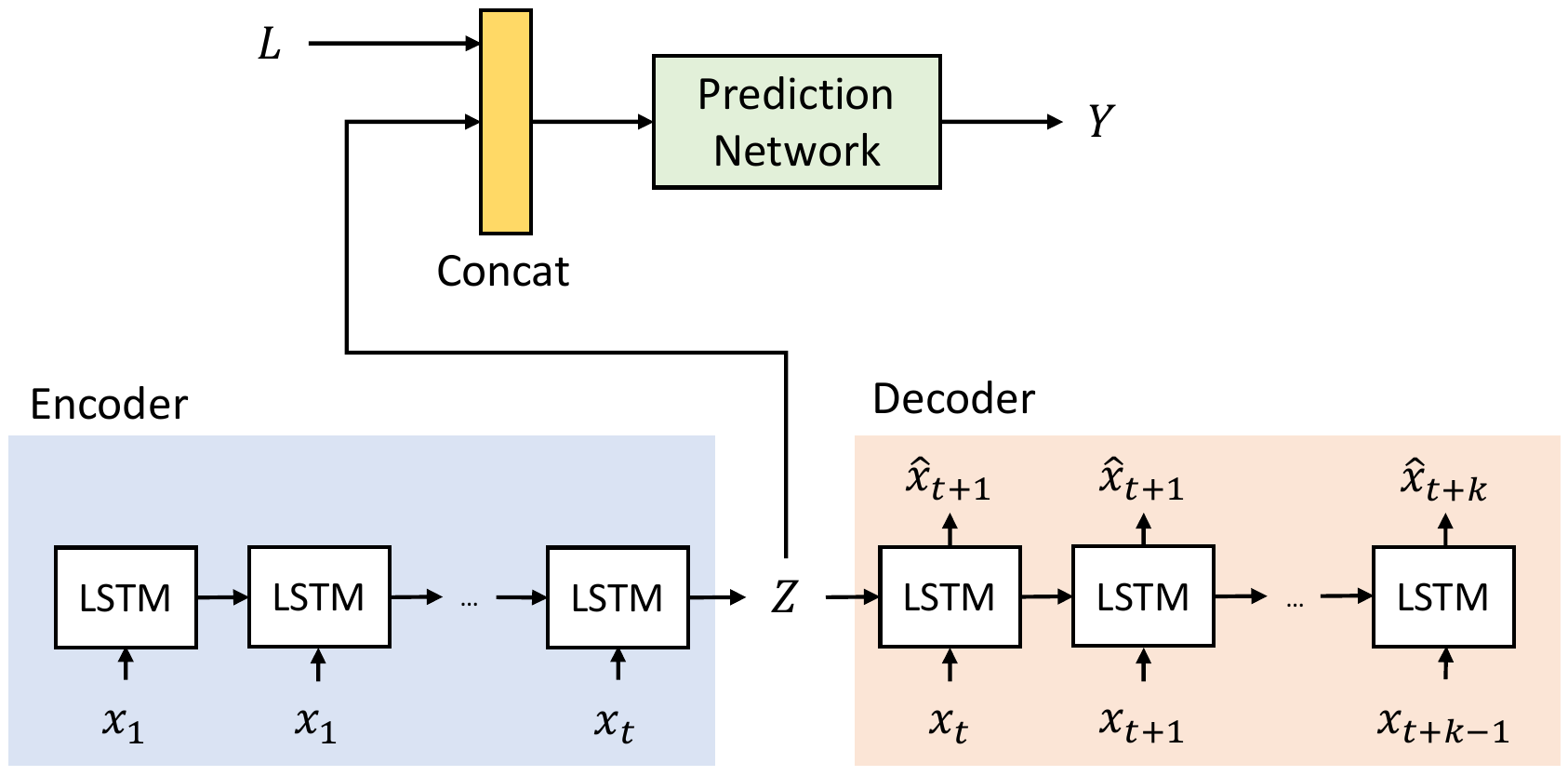}
	\caption{\SCHEDULER's hybrid Bayesian model for the dynamic pre-warmed function container pool, consisting of a LSTM encoder-decoder and a prediction network. } 
    \label{fig:hybrid_model}
\end{figure}

Classic timeseries prediction models (e.g., exponential smoothing, ARIMA models, Theta method) usually require manual tuning to configure the model and uncertainty parameters~\cite{hyndman2008forecast}. Moreover, it is difficult to incorporate external features into these models, which can be impactful to accuracy. Long Short Term Memory (LSTM) models~\cite{hochreiter1997lstm} have also gained popularity in timeseries prediction. LSTM can capture long-term sequential dependencies in the data and outperform traditional methods~\cite{laptev2017time}. However, conventional LSTM models cannot easily embed non-temporal external features or incorporate noise and uncertainty into their predictions, which is important for handling fluctuating workloads.

To overcome these problems and achieve generalizable and scalable prediction, we build a hybrid Bayesian neural network model. The novelty of our Bayesian model is twofold. First, it can utilize external features, such as time of day, to forecast function invocations. Second, it takes system noise into account when making predictions, allowing it to  provide reliable uncertainty estimation, which is critical for fluctuating workloads.
As shown in Fig.~\ref{fig:hybrid_model}, the model consists of two parts: (i) the Long Short-Term Memory (LSTM) encoder-decoder, which serves as a feature-detection blackbox that extracts a latent variable from the input timeseries; and (ii) the prediction network, which infers the invocation pattern in the next time window using the latent variable and external features. We use Monte Carlo (MC) dropout~\cite{gal2016dropout} to approximate Bayesian inference and quantify the prediction uncertainty. 

\noindent{\bf{LSTM encoder-decoder:}} 
Before training the prediction model, we first construct and train the LSTM encoder-decoder to extract latent features from a serverless trace, which contains information of the historical invocation patterns. The LSTM encoder-decoder consists of two LSTMs modules. The encoder processes the input workload sequence ($X=\{{x_1}, {x_2}, ..., {x_t}\}$), and generates the latent variable ($Z$), which summarizes its information. The decoder uses the extracted latent variable to produce the output workload sequence for the upcoming $k$ windows $\{{x_{t+1}}, {x_{t+2}}, ..., {x_{t+k}}\}$.
The LSTM encoder-decoder is constructed using stacked LSTM cells with two layers. The encoder and decoder have 64 and 16 features in the hidden states respectively; the network's configuration is discussed below. 

\noindent{\bf{Prediction network:}} 
After training the LSTM encoder-decoder, we use the LSTM encoder as an automatic feature-extraction blackbox. The last hidden state of the encoder is the latent variable $Z$. Then, we train a prediction network to forecast the number of active containers ($Y$) in the next time window, using $Z$ as features. To further increase the prediction accuracy, we concatenate the external feature vector ($L$) with $Z$, then feed it into the prediction network. We build the prediction network using a multi-layer perceptron, which consists of tanh activation functions and three fully connected layers.
The model parameters of the LSTM and prediction network are selected based on the validation accuracy.
\zz{I added a short explanation on the model parameter selection. Yes I have separated training and testing dataset. This explanation is basically the same as Sinan. We use variational dropout to avoid over-fitting}

\noindent{\bf{Bayesian inference:}} Incorporating noise and uncertainty into the model is essential for accurate timeseries forecasting under fluctuating load. To enable this, we leverage approximate Bayesian inference.
Due to its simplicity, generality and scalability, we use MC dropout~\cite{gal2016dropout} to approximate Bayesian neural networks and achieve epistemic uncertainty estimates, rather than training a deterministic model. We apply variational dropout to the encoder~\cite{gal2016rnn}, and regular dropout to the prediction network. By applying stochastic dropouts to each hidden layer of the encoder and prediction network, we can obtain the predictive mean and variance through $T$ forward passes using different samples of model weights ($\{W_t\}^T_{t=1}$). 




\subsection{Prediction-Based Container Pool Manager} 

\SCHEDULER adjusts the number of pre-warmed containers for the next time window based on model predictions, by creating warm containers in advance to accommodate incoming invocations, and shutting down idle containers in time to save resources.
The adjustment interval of the container pool is 1 minute, which is long enough to hide the container instantiation overhead, and is the typical time-scale for container keep-alive times in production FaaS platforms~\cite{shahrad2020serverless}.
The latency of the prediction model is below 10ms, which is negligible compared to the adjustment interval of the container pool. 

%% file: Design_BO.tex
\section{Optimizing Per-Function Resources}
\label{sec:bo}

Cold starts are not the sole reason for performance degradation in FaaS platforms. It is also critical to ensure that the resources allocated to each function are appropriate. The pre-warmed container pool manager ensures that the majority of function invocations are handled by warm containers, which simplifies function-level resource allocation, narrowing it down to only considering the warm-start performance behavior of serverless workflows. 

\SCHEDULER needs to consider the diverse resource requirements of each function across execution stages. Manually deriving an analytical performance model for a variety of applications is difficult. On the other hand, exhaustively searching the entire configuration space is time consuming and expensive, since the total number of available configurations grows exponentially with the number of stages in a workflow. Moreover, each configuration needs to be profiled multiple times to get around the noise in FaaS platforms. 

Rather than relying on manually-derived analytical models or exhaustive profiling, \SCHEDULER uses Bayesian Optimization (BO), a data-driven approach, to learn the mapping from resource configurations to performance and cost. BO has been effective in black-box resource optimization for long-running cloud workloads~\cite{patel2020clite,roy2021satori,alipourfard2017cherrypick}, where application behaviors are not known to the cloud provider in advance. However, previous BO-based resource managers did not take noise and uncertainty into account, leading to increased search time and cost, and degraded performance.
\SCHEDULER's container resource manager leverages an improved Bayesian Optimization (BO) approach that considers noise and uncertainty and is robust to biased observations and data outliers, resulting in fast convergence and lower search overheads. \SCHEDULER also exploits the scalability of serverless workloads to accelerate exploration by enabling batch sampling, rather than using individual samples as in previous work. 

We first describe the BO algorithm workflow, and then discuss the challenges that prevent conventional BO from being robust to noise in FaaS platforms. Finally, we discuss \SCHEDULER's customized BO that overcomes these challenges. 

\subsection{Bayesian Optimization Workflow} 
\label{subsec:bo}
\noindent{\bf{Problem formulation:}} 
Formally, for a multi-stage serverless application, we want to find the resources ($c$) that minimize execution cost ($f$), while satisfying the end-to-end QoS ($\lambda$); the formula is shown in Eq.~\ref{eq:bo_opt}. The resource configuration includes the CPU, memory, and concurrency settings for all functions in the application, consistent with the interface of major FaaS providers~\cite{awslambda, googlefunction, azurefuntion}. 
The execution cost is linear to the CPU and memory time, consistent with cost models in production serverless platforms~\cite{awslambda,googlefunction,azurefuntion}. The optimization objective is: 
\vspace{-0.08in}

\begin{equation}
    \min_{c} f(c) \ \ \text{subjects to} \ \ \ell(c) \leq \lambda
    \label{eq:bo_opt}
\end{equation}

\vspace{-0.08in}
$f(c)$ and $\ell(c)$ are black-box functions, whose values (cost and execution time respectively) can be observed by sampling resource configuration $c$. Collecting more samples increases the probability of finding a good configuration, at the cost of increased exploration overheads. However, the search process is under both time and budget constraints, as shown in Eq.~\ref{eq:bo_training}, in which $T_{budget}$ denotes the budget towards sampling resource configurations $\{{c_{1}}, {c_{2}}, ..., {c_{k}}\}$, and $T_{time}$ indicates the time constraint for the exploration process.

\vspace{-0.08in}
\begin{equation}
    \sum_{k=1}^{K}f(c_k) \leq T_{budget} \quad \text{and} \quad  \sum_{k=1}^{K}\ell(c_k) \leq T_{time}
    \label{eq:bo_training}
\end{equation}
\vspace{-0.08in}
 


\begin{figure}[t]
     \centering 
     \includegraphics[width=0.90\columnwidth]{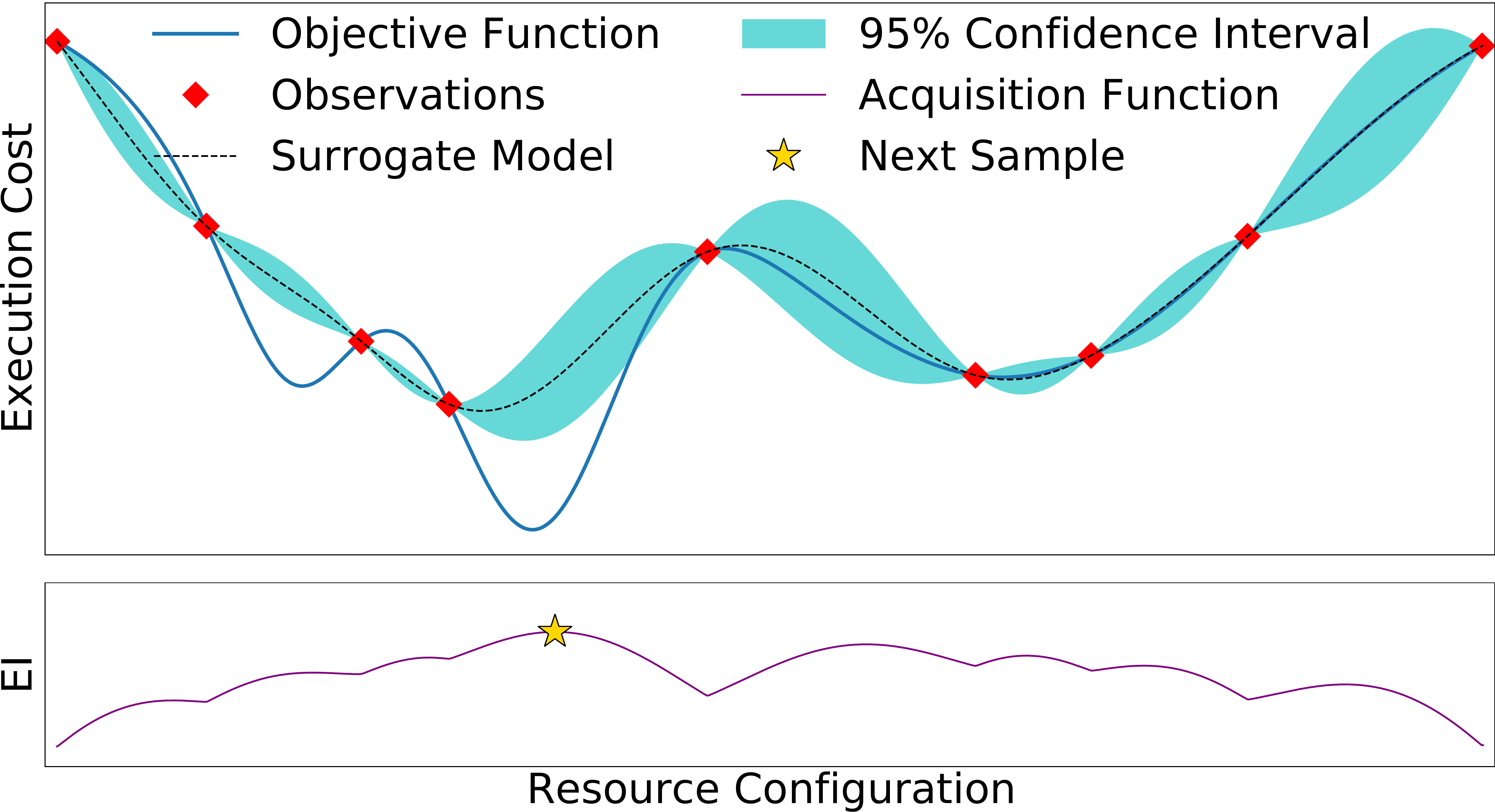}
 	\caption{Iterative process in Bayesian Optimization (BO). }
     \label{fig:bo_visual}
\end{figure}

\noindent{\bf{Bayesian optimization:}}
BO relies on two key components. First, BO relies on a model that captures the relationship between input and objective function to drive the optimization process, a model commonly referred to as surrogate model in BO literature~\cite{shahriari2016taking}. Second, BO leverages acquisition functions that determine the next data point to be sampled based on the predictions of the surrogate model.
As shown in Fig.~\ref{fig:bo_visual}, the algorithm proceeds iteratively and in each epoch, the surrogate model is updated with the data (resource configuration and corresponding performance metrics) sampled in the previous epoch, and the acquisition function leverages the updated surrogate model to determine the next data point (candidate resource configuration) to be sampled in the current epoch.

\subsection{Challenges for Conventional BO} 
\label{subsec:bo_challenges}

Conventional BO-based resource managers can suffer from increased search time and cost, and degraded performance due to the following challenges: 

\begin{itemize}[leftmargin=*]
\item {\bf Cloud noise:} Previous BO-based resource managers assume a noiseless setting~\cite{patel2020clite}\cite{roy2021satori}\cite{shahriari2016taking}.
However, the cloud is a noisy environment. For example, resource interference and workload fluctuation, can exacerbate performance unpredictability, and result in biased observations of workload performance. Serverless applications can also suffer from interference, leading to misleading observations (outliers) in BO's sampling process. In this case, the naive BO workflow would suffer from model misspecifications caused by outliers, and the GP models would fail to characterize the performance of the workflow. 

\item {\bf QoS constraint:} Adding black-box inequality constraints like QoS constraints to BO is challenging~\cite{shahriari2016taking}. Prior BO-based resource managers~\cite{patel2020clite,roy2021satori} rely on manually crafted objective functions with a penalty term that is triggered upon QoS violation, to guide the sampling process. However, manually crafted objective functions lack the flexibility to capture the behavior of complex serverless workflows and can lead to slow convergence and performance degradation.

\item {\bf Batch sampling:} Conventional BO samples and evaluates one configuration at a time~\cite{shahriari2016taking}, limiting the speed of convergence. For serverless applications, if we take advantage of the scalability of serverless by sampling multiple configurations at a time, the exploration can be greatly accelerated, improving the resource savings. 
\end{itemize}


\subsection{Customized Bayesian Optimization} 
\label{subsec:acquisition}

We propose a customized BO which addresses the challenges above.
The algorithmic novelty of this customized BO is threefold
\begin{itemize}[leftmargin=*]

\item {
First, different from previous approaches that ignore or underestimate cloud noises, Aquatope takes noise and uncertainty into account by design, when searching for a near-optimal resource configuration. To capture various forms of cloud noise, Aquatope divides noises into two categories: one is inherent noise, which can be approximated well by a normal distribution; the other is irregular noise which does not follow a normal distribution. The latter includes noise caused by resource contention or networking instability. We refer to the first type of noise as \textit{Gaussian noise} and to the second as \textit{non-Gaussian noise}. Aquatope uses noise-aware surrogate models and acquisition functions to account for Gaussian noise, and builds diagnostic models to prune the non-Gaussian data outliers.
}

\item {
Second, Aquatope effectively incorporates end-to-end QoS constraints into BO. Unlike conventional BO that relies on a manually crafted objective function with a reactive penalty term that is triggered when a QoS violation occurs, Aquatope takes a proactive approach by building a surrogate model that predicts end-to-end performance, and uses the predictions of the model to filter candidate configurations that may violate QoS.
}

\item {
Finally, instead of sampling one configuration at a time, Aquatope employs batch sampling with customized acquisition functions, substantially reducing the exploration time, without sacrificing the quality of the selected resource allocation configuration. 
}

\end{itemize}


\noindent{\bf{Customized  surrogate  models:}}
\SCHEDULER uses Gaussian process (GP)~\cite{rasmussen2005gaussian} as the surrogate model. GP is a suitable surrogate model for resource exploration for several reasons. GP is non-parametric and does not make any assumptions over the target black-box function and is thus flexible enough to capture the relationship between resources and performance. GP is also computationally tractable and can be evaluated and updated cheaply and often~\cite{brochu2010tutorial}. Finally, GP can provide a measure of uncertainty for the predictions of unsampled data points and naturally captures Gaussian noise.
Specifically, \SCHEDULER uses fixed-noise GP models with Mat{\'e}rn(5/2) as the covariance kernel~\cite{rasmussen2005gaussian} to model the Gaussian noise. 

More importantly, instead of combining the cost and performance targets with a manually crafted objective function~\cite{patel2020clite, roy2021satori} and building a single GP model for it, \SCHEDULER builds independent GP models for the cost target $f$ and the QoS constraint $\ell$. The intuition for separating the two is to allow the GP models to converge faster and more accurately. The cost GP model captures the cost reduction for an unsampled resource configuration, and the performance GP model  narrows down the search space, by discerning the regions more likely to be feasible (i.e., satisfy QoS). 

\begin{figure}[t]
    \centering 
    \includegraphics[width=\columnwidth]{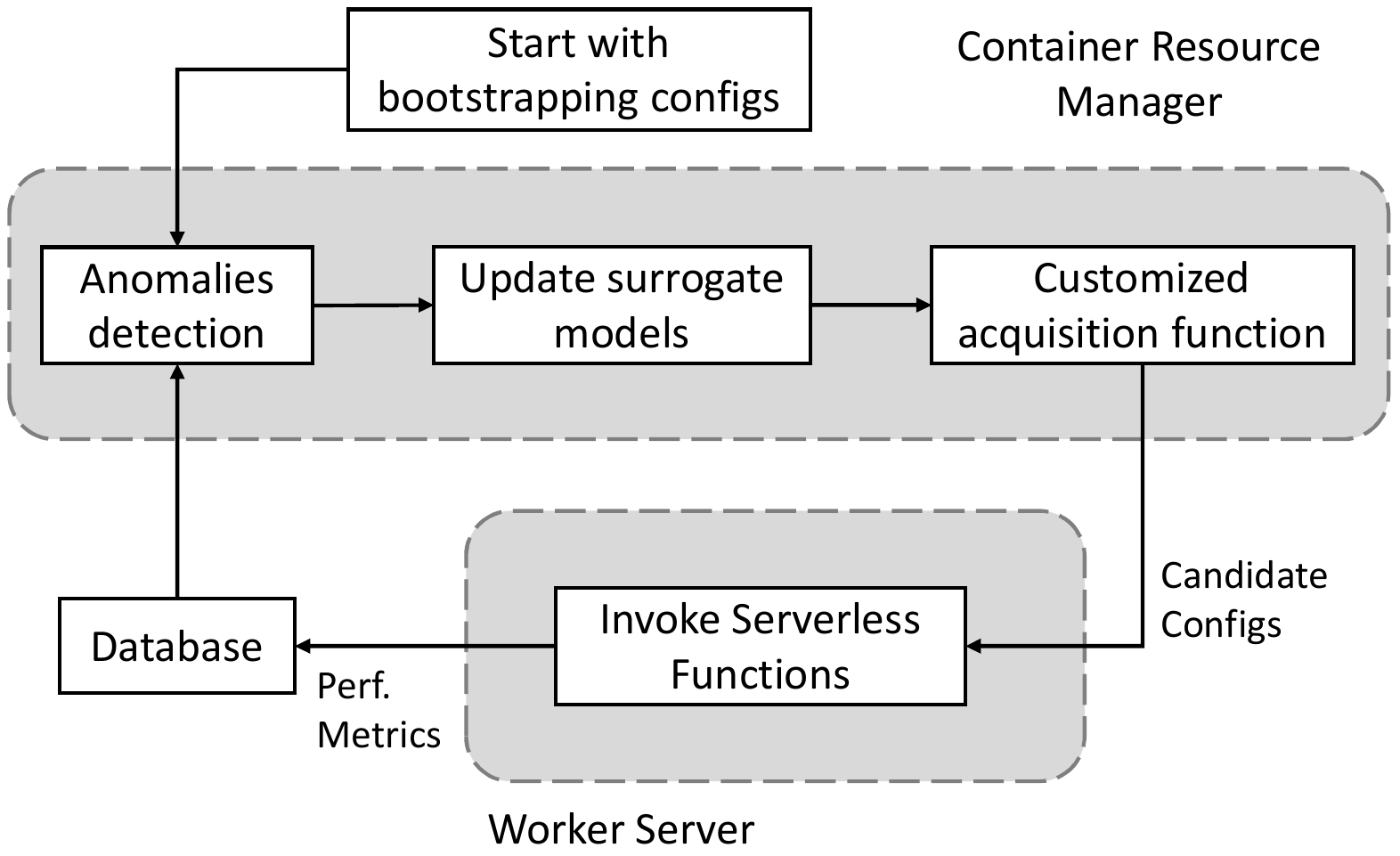}
	\caption{Workflow of \SCHEDULER's container resource manager.}
    \label{fig:bo_workflow}
    \vspace{-0.03in}
\end{figure}

\noindent{\bf{Customized acquisition function:}}
\SCHEDULER uses customized acquisition functions to select the next batch of candidate configurations, maximizing expectation of improvement (cost reduction) over the current best observation.
The classic expected improvement (EI) acquisition function~\cite{snoek2012practical} provides a reasonable balance between exploration and exploitation at a low computation cost. However, EI selects one candidate in each iteration and assumes noiseless observations. Instead, we leverage recent advances in BO to use constrained noisy expected improvement (NEI) with quasi-Monte Carlo integration (QMC)~\cite{letham2019constrained}. NEI takes Gaussian observation noise into consideration and does not assume the best observation is known, which would require noiseless observations.

We use the method in~\cite{gardner2014bayesian} to multiply NEI of reducing cost with the probability of satisfying QoS, which is derived from the performance GP model, to obtain the constrained NEI. The constrained NEI helps \SCHEDULER to focus on the feasible configuration space, where QoS can be met. QMC provides an approximation of constrained NEI and its gradient, which do not have analytic expressions, and enables batch optimization by iteratively maximizing NEI integrated over pending unobserved samples. We use a batch size of 3, which speeds up the search without sacrificing quality.

\noindent{\bf{Anomaly detection:}} We refer to data outliers from non-Gaussian noise as anomalies.
\SCHEDULER builds diagnostic models to prune anomalies in the sampling process. For each sampled configuration, we create a diagnostic GP model using data points other than the one under evaluation. The diagnostic GP model computes the predictive mean and confidence interval to identify a possible anomaly. If the observed value of that configuration falls outside the 95\% predictive confidence interval, it is labeled as an anomaly. We evaluate all observed configurations and add potential anomalies to the list. 

\noindent{\bf{Batch evaluation:}}
After obtaining a batch of candidate configurations, \SCHEDULER sends requests to the pre-warmed container pool to launch the serverless workflow, to ensure warm starts. Then it profiles all candidate configurations in parallel and evaluates their performance. We use both QoS-preserving and QoS-violating sample observations to update the surrogate models, because 
QoS-violating configurations help the GP models to identify which regions are more likely to meet QoS  without actually sampling them. 

\noindent{\bf{Putting it all together:}}
The complete workflow of the customized BO engine is shown in Fig.~\ref{fig:bo_workflow}. The BO engine starts with a few randomly sampled configurations to warm up the surrogate models. Then the BO engine proceeds iteratively. In each iteration, the BO engine uses the customized acquisition functions to select a batch of candidate configurations to sample that are likely to preserve QoS. When the sampling finishes and performance metrics are retrieved, the observed performance metrics are first sent to the anomalies detection engine to filter misleading observations, which are then used to update both the performance and cost surrogate models.


\noindent{\bf{Incremental retraining:}} The anomaly detection mechanism also allows \SCHEDULER to detect changes in the performance behavior of serverless workflows, when the observed performance metrics deviate from the model predictions. These deviations can be caused by changes in the input workload, function updates, etc. In this scenario, \SCHEDULER performs incremental retraining, and updates the model by collecting new samples using a sliding window, and gradually adapts to changes in the application behavior.

%% file: Implementation.tex
\section{System Implementation}
\label{sec:implementation}

\begin{figure}[t]
    \centering 
    \includegraphics[width=\columnwidth]{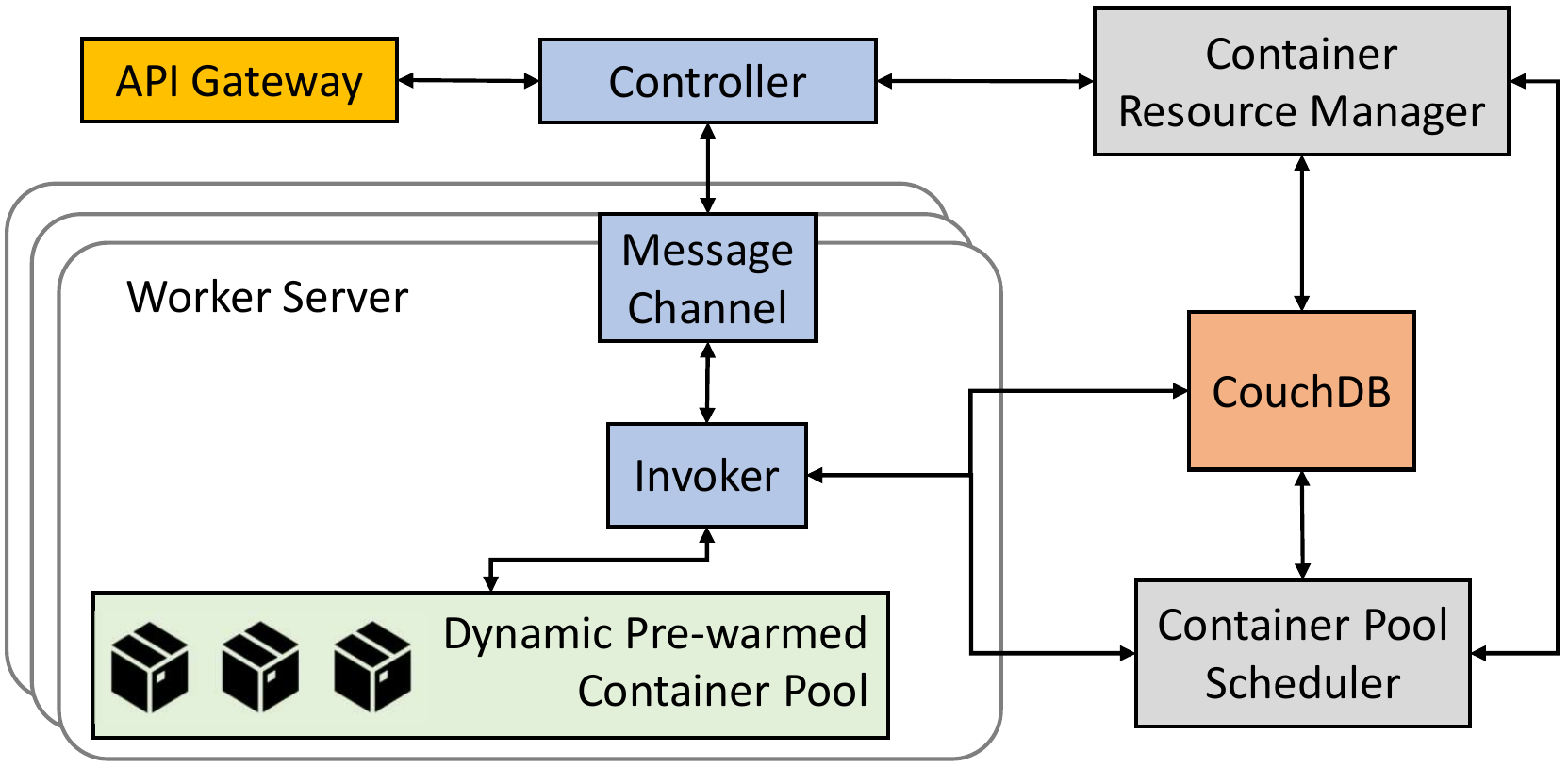}
	\caption{Architecture of \SCHEDULER's implementation. }
    \label{fig:system_arch}
\end{figure}

\SCHEDULER is built over Apache OpenWhisk~\cite{openwhisk}; a widely-used open-source FaaS platform that powers IBM’s Cloud Functions~\cite{ibmfunction}. Fig.~\ref{fig:system_arch} shows \SCHEDULER's implementation. 

\noindent{\bf{OpenWhisk architecture:}}
The API gateway of OpenWhisk is implemented with NGINX~\cite{nginx}. The backend of OpenWhisk consists of controllers and invokers that scale horizontally, with one invoker deployed per worker server. Function invocations are forwarded to a controller, which chooses an invoker to execute the invocation by considering invoker capacity and execution history. The invocation is sent to the invoker through a message channel implemented with Kafka~\cite{kafka}. Function implementations, invocation histories, execution results, and statistics are stored in CouchDB~\cite{couchdb}. 

\noindent{\bf{Resource scheduling:}} 
By default, OpenWhisk allocates a relative share of CPU proportional to the amount of memory provisioned for each function container. To implement \SCHEDULER, we modified the resource scheduling mechanism of OpenWhisk to decouple CPU and memory resource allocations, and support CPU-limit-based resource scheduling. 

\noindent{\bf{Dynamic pre-warmed container pool:}}
Similar to AWS Lambda's provisioned concurrency~\cite{provisioned}, OpenWhisk's invoker maintains a pool of pre-warmed containers (\textit{stem cell}) for heavily-used functions. By default, the configuration of the pre-warmed container pool is static and pre-defined, and all worker servers share the same configuration. We modify the controller and invoker to support dynamic adjustment of the pre-warmed container pool, making it worker-server specific, and configured via the controller for all managed invokers (or via the invoker directly). The load balancer in the controller is aware of the pre-warmed containers and routes function invocation requests to the supporting invokers accordingly. 

\noindent{\bf{Container pool scheduler:}} 
\SCHEDULER runs an independent service to control the pre-warmed containers. It fetches metadata for the serverless applications requiring pre-warmed resources, and their invocation histories from CouchDB. For each application, the scheduler trains the prediction model, and uses it to adjust the dynamic pre-warmed container pool. The hybrid Bayesian NN is implemented with PyTorch~\cite{pytorch}. The scheduler makes decisions in each time interval and sends the updated container pool configurations to the invokers. 

\noindent{\bf{Container resource manager: }} \SCHEDULER aims to find a near-optimal configuration for a serverless application. When a new application is registered, \SCHEDULER obtains its metadata and QoS from CouchDB, and starts the optimization process. The GP models are implemented using GPyTorch~\cite{gardner2018gpytorch} and the optimization workflow is implemented in BoTorch~\cite{balandat2020botorch}. The engine samples the candidate resource configurations on the worker servers. The execution results and performance metrics are fetched from CouchDB. After selecting a near-optimal configuration, the engine sends messages to the controller to update the configuration of the application.

%% file: Methodology.tex
\section{Methodology}
\label{sec:methodology}

\begin{figure}[t]
    \centering 
    \includegraphics[width=1.00\columnwidth]{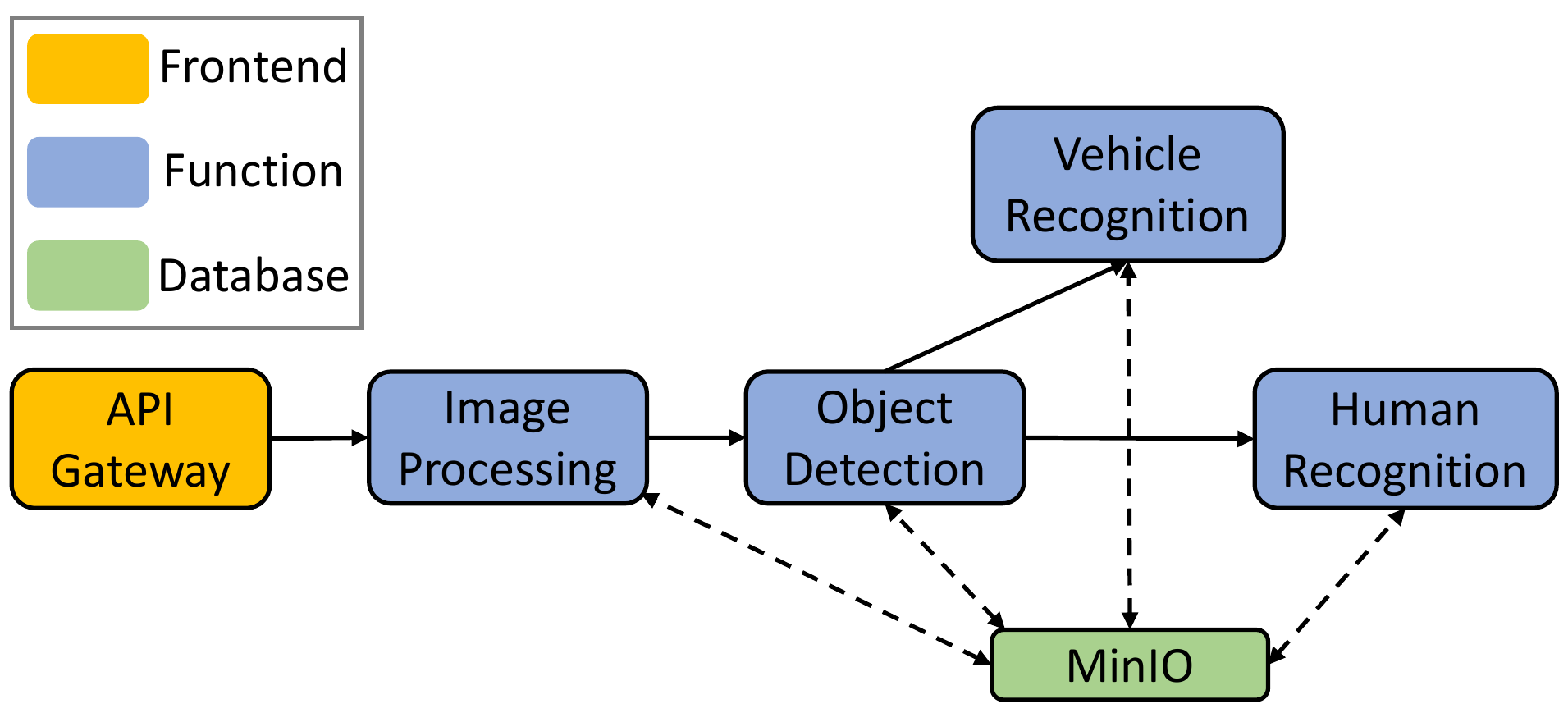}
	\caption{ML pipeline architecture.}
    \label{fig:benchmark_ml_pipeline}
\end{figure}

\begin{figure}[t]
    \centering 
    \includegraphics[width=1.00\columnwidth]{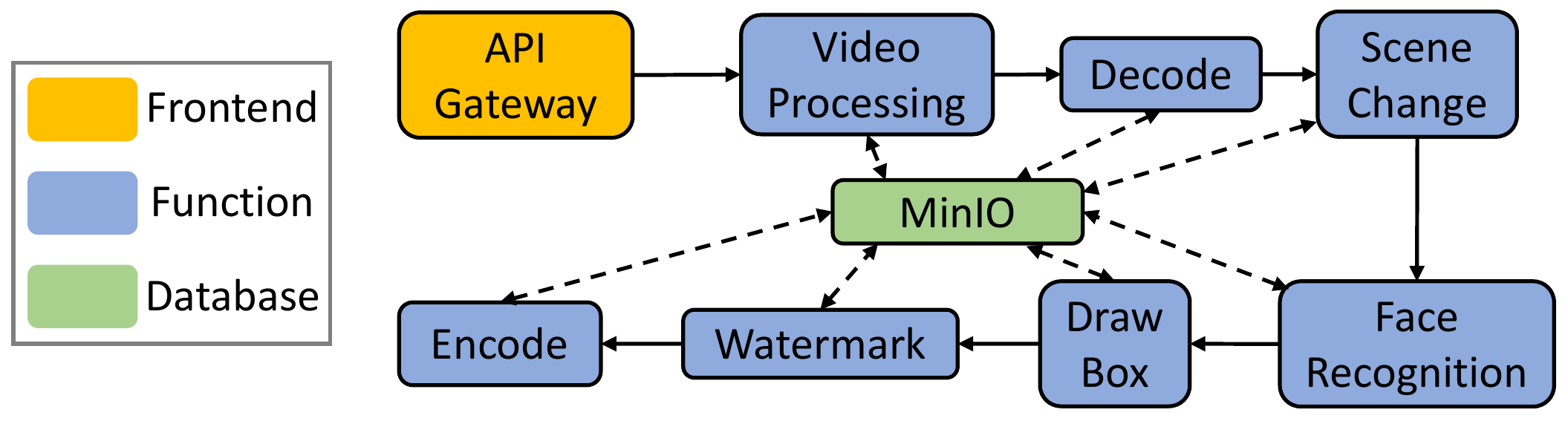}
	\caption{Video processing framework architecture.}
    \label{fig:benchmark_video}
\end{figure}

\subsection{Applications}

\noindent{\bf{Generic function workflows:}} 
We first implement several generic function workflows using the Apache OpenWhisk Composer~\cite{composer}, to combine multiple synthetic serverless functions into multi-stage workflows. We create a function generator to synthesize configurable resource-intensive functions that emulate varying CPU and memory workloads. 
We generate two workflows which are often present in multi-stage workflows: Chain and Fan-out/Fan-in.
In Chain, a sequence of functions executes in a specific order. The output of one function is fed to its downstream function. In Fan-in/Fan-out, the workflow executes multiple functions in parallel, and only returns when the last child completes after some aggregation. 


\begin{figure}[t]
    \centering
    \includegraphics[width=1.00\columnwidth]{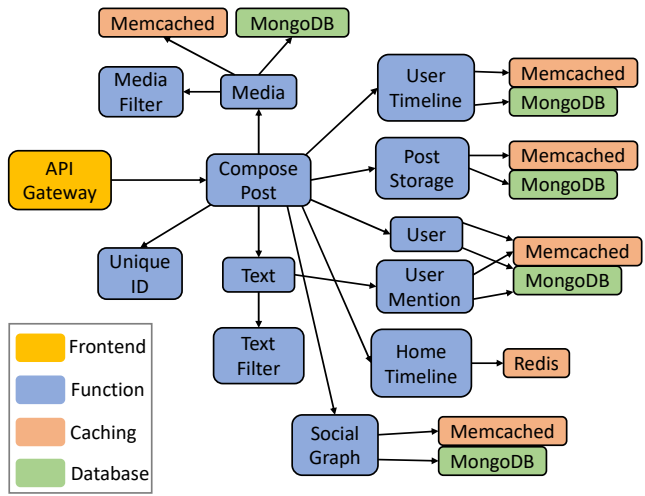}
	\caption{Serverless social network architecture~\cite{gan2019asplos}.}
    \label{fig:benchmark_social}
\end{figure}

\vspace{0.03in}
\noindent{\bf{ML pipeline:}}
We implement an ML pipeline which serves as the backend of a parking lot security system, trying to recognize humans and vehicles. Fig.~\ref{fig:benchmark_ml_pipeline} shows the architecture of the ML pipeline. The image recorded by the parking lot security camera is uploaded to the object store and triggers the ML pipeline. The pipeline performs object detection~\cite{jian2020gluoncv} and the labeled images are uploaded to the object store, with vehicle and human recognition being invoked in parallel.

\vspace{0.03in}
\noindent{\bf{Video processing framework:}}
We implement a serverless video processing framework similar to Sprocket~\cite{ao2019sprocket}. The input video URLs are fetched and decoded into fixed-length frames. Then different function pipelines are invoked and the video chunks are processed in parallel. Fig.~\ref{fig:benchmark_video} shows the framework's architecture. We use MinIO~\cite{minio} as the ephemeral storage for the video frames for each stage. 

\vspace{0.03in}
\noindent{\bf{Social network:}}
We use a serverless implementation of the broadcast-style Social Network in DeathStarBench~\cite{gan2019asplos}. Fig.~\ref{fig:benchmark_social} shows the architecture of the serverless implementation of the service. Users can create posts embedded with text, media, links, and tags, which are then broadcast to all their followers. The texts and images uploaded by users go through the \textit{text-filter} and \textit{image-filter} functions. Contents violating the service’s ethical guidelines are rejected. Users can also read posts on their timelines. The backend uses Memcached and Redis for caching, and MongoDB for persistent storage.
We use the socfb-Reed98 Facebook network dataset~\cite{ryan2015network} as the social graph, with 962 users and 18.8K follow relationships.

\subsection{Workload Generation}
We use the Locust~\cite{locust} load generator to emulate real user traffic. We generate custom-shaped loads based on scaled-down invocation pattern traces from the Azure Function Dataset~\cite{shahrad2020serverless}. Since the Azure dataset does not contain traces of Azure Durable Functions (the workflow engine for composing function logic)~\cite{azuredurable}, we use the function invocation traces to emulate workflow invocation patterns. Within each one-minute interval provided in the trace, we use a Poisson process to generate workflow invocation traffic with an exponential distribution of inter-arrival times. We scale the invocation rate proportionally so that the maximum CPU utilization in the cluster does not exceed 70\%, which is in accordance with the CPU utilization in the Google and Alibaba production clusters~\cite{shan2018lego}, and the Azure Function cluster~\cite{zhang2021harvested}. This workload generation method is consistent with the methodology in~\cite{shahrad2020serverless}. The load generators and functions are never physically co-located on a server. 

\subsection{Server Cluster}

We deploy \SCHEDULER to a dedicated local cluster with five, 2-socket, 40-core servers using Intel x86 Xeon E5s with 128GB RAM each, and two 2-socket, 88-core servers using Intel Gold 6152 processors with 188GB RAM each. Each server is connected to a 40Gbps ToR switch over 10Gbe NICs. All machines run Ubuntu 18.04.3 LTS. We use one of the 40-core servers to host the controller, API gateway, CouchDB and other system components, including \SCHEDULER. Each of the remaining servers hosts an invoker and maintains a dynamic pre-warmed container pool to run the functions using Docker. 

\subsection{Comparison Baselines}
We compare \SCHEDULER with multiple strategies that mitigate cold starts and optimize resource allocations.
In terms of reducing cold starts, we compare against (1) the fixed keep-alive policy used by most FaaS providers~\cite{awslambda, azurefuntion}; (2) Apache OpenWhisk's reactive-stem-cell policy~\cite{openwhiskreactive}, which enables autoscaling for pre-warmed containers; (3) FaaSCache's container eviction and dynamic auto-scaling policy~\cite{fuerst2021faascache}; (4) histogram-based container keep-alive policy in~\cite{shahrad2020serverless}, which uses historical function inter-arrival time to dynamically adjust the keep-alive time; (5) Icebreaker~\cite{roy2022icebreaker}, which uses Fourier Transformation to predict and pre-warm function containers based on historical invocation patterns. 

For resource management, we compare \SCHEDULER with (a) autoscaling techniques~\cite{saha2018emars, suresh2020ensure}, that dynamically adjust a container's CPU and memory to match the function's latency requirements; (b) a random-search-based tuning system~\cite{herodotou2011starfish};
and (c) CLITE~\cite{patel2020clite}, which uses a BO-driven approach to search for a near-optimal resource configuration that minimizes the cost while satisfying QoS. We modify CLITE's score function to make it applicable to multi-stage serverless applications. Details are provided in Section~\ref{subsec:eval_bo}.

%% file: Evaluation.tex
\section{Evaluation}
\label{sec:evaluation}

We first evaluate \SCHEDULER's two key components ( dynamic pre-warmed container pool and container
resource manager) separately, and then perform an end-to-end evaluation that includes both components. 

\subsection{Dynamic Pre-warmed Container Pool}
\label{subsec:prewarm}

\noindent{\bf{Prediction model accuracy:}}
We first evaluate the accuracy of the hybrid Bayesian NN used to predict the number of pre-warmed containers in \SCHEDULER, by measuring its average accuracy across different serverless workflows and invocation patterns.
We also compare \SCHEDULER's model with three alternatives: (1) fixed Keep-Alive:  A na{\"i}ve model that uses the number of invoked containers in the last time window as the prediction for the next. (2) ARIMA~\cite{hyndman2008forecast}: Auto-Regressive Integrated Moving Average, a classic timeseries  prediction model used in Microsoft Azure's ``Serverless in the Wild'' system~\cite{shahrad2020serverless}, and (3) LSTM~\cite{hochreiter1997lstm}: a vanilla LSTM model with similar configuration as our hybrid model, but without considering external features, such as time  of  day/week and  function types, or taking uncertainty into account. We use the same training dataset for all systems and evaluate performance on a separate test dataset.


Table~\ref{tbl:model_accuracy} shows the Symmetric Mean Absolute Percentage Error (SMAPE) of the four models across all workflows in terms of pre-warmed vs. required containers, a widely used metric in time series prediction~\cite{zhu2017deep}. 
\SCHEDULER's hybrid model significantly outperforms all other alternatives, with a 40\% reduction in prediction error compared to the second best model, the vanilla LSTM.
Our proposed Bayesian NN outperforms fixed Keep-Alive and ARIMA because these simple analytical models do not fully capture the dynamic invocation pattern, and it outperforms the vanilla LSTM because it uses information-rich external features as input, and also takes into account cloud noise and uncertainty when making predictions.

\begin{figure}[t]
    \subfloat[Function cold starts.\label{subfig:eval_cold_start}]
    {
      \includegraphics[width=0.426\columnwidth]{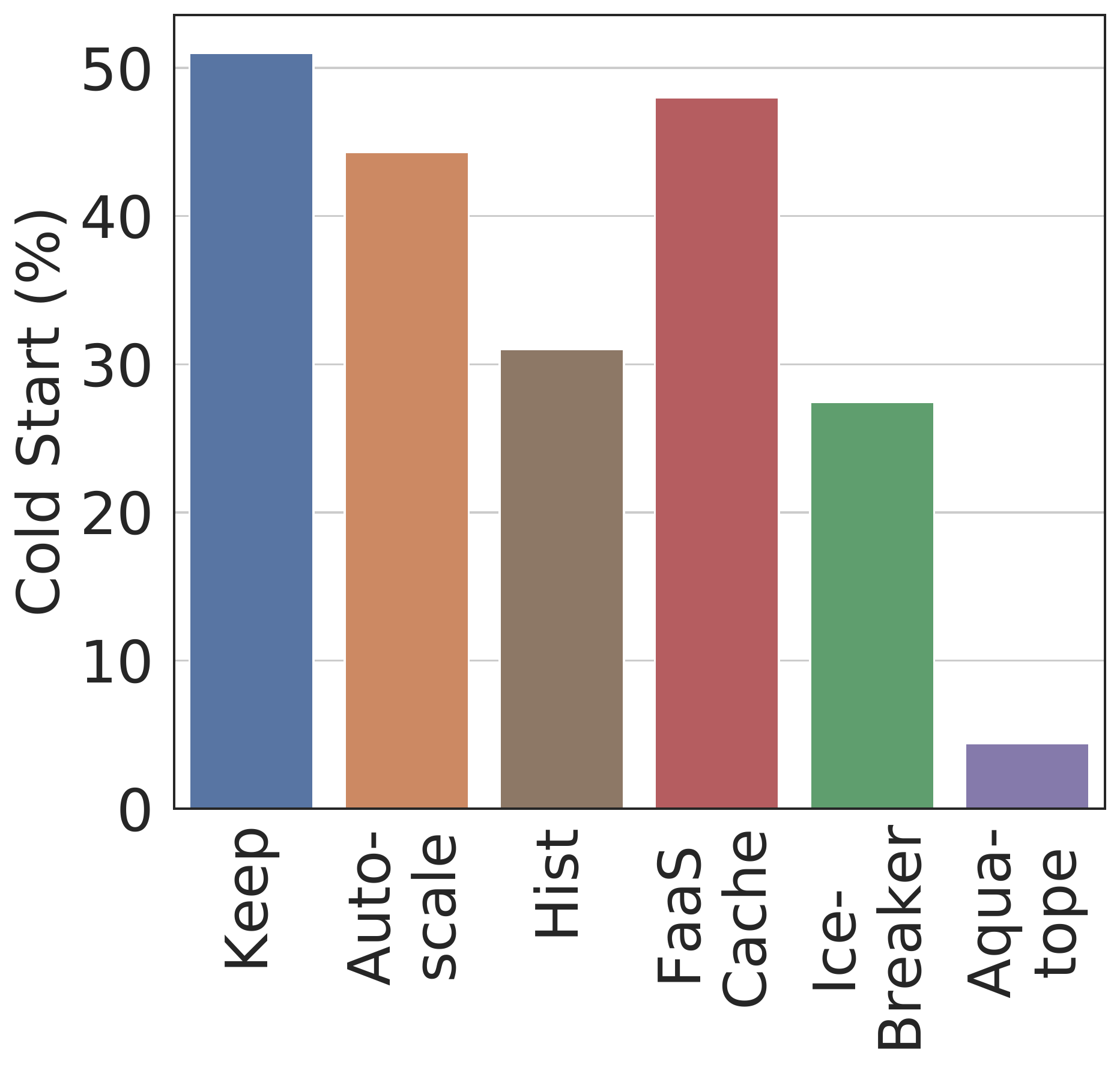}
    }
    \hspace{0.2cm}
    \subfloat[Provisioned memory time.\label{subfig:eval_wasted_memory}]
    {
      \includegraphics[width=0.44\columnwidth]{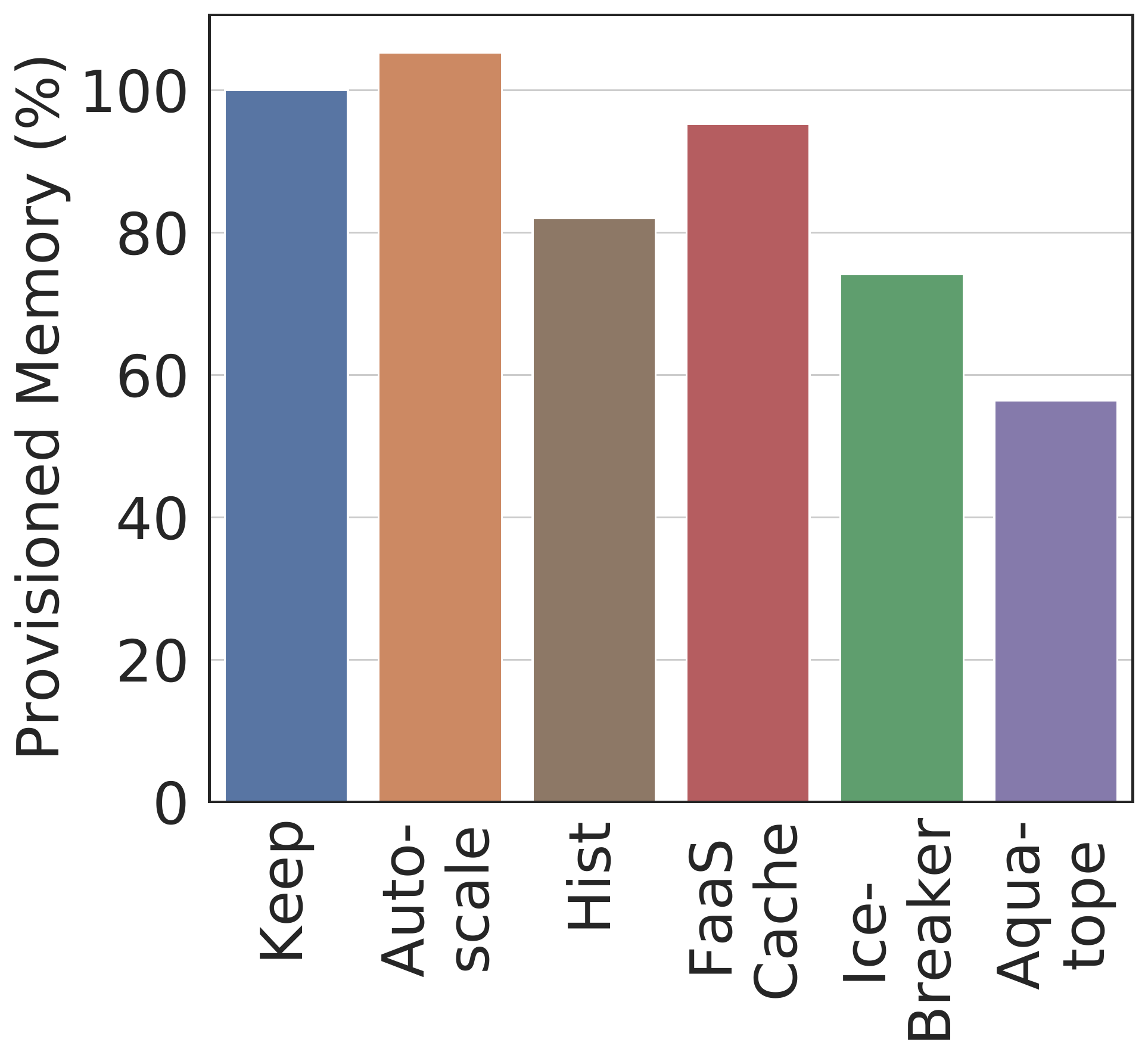}
    }
\caption{\SCHEDULER's dynamic pre-warmed container pool outperforms other empirical and data-driven approaches. }
\label{fig:eval_cold_start}
\end{figure}

\vspace{0.03in}
\noindent{\bf{Eliminating cold starts:}}
We now evaluate the cold start elimination approach in \SCHEDULER compared to previous work. The results are shown in Fig.~\ref{subfig:eval_cold_start}.

\begin{table}[t]
\caption{Prediction accuracy measured in SMAPE. } 
\centering
\label{tbl:model_accuracy}
\begin{adjustbox}{width=1.0\linewidth}
\begin{tabular}{c||cccc}
\toprule
\textbf{Prediction}& \multicolumn{4}{c}{\bf Prediction Models} \\
\textbf{Error} & Fixed Keep-Alive & ARIMA & LSTM & {\bf \SCHEDULER} \\ \toprule
\textbf{SMAPE} & 24.5\% & 18.6\%  & 9.5\% & 5.7\% \\ \bottomrule
\end{tabular}
\end{adjustbox}
\end{table}



The fixed \textit{Keep-Alive} policy keeps containers alive for another 10 minutes after executing the last invocation, and the resulting the cold start rate is 51\%. The autoscaling policy~\cite{openwhiskreactive, lambdascaling}
adjusts the number of pre-warmed containers based on utilization, and achieves cold start rate of 44\%.
However, autoscaling relies on reactive feedback control, and cannot adjust the containers fast enough, when load fluctuates rapidly.  FaaSCache~\cite{fuerst2021faascache} performs similarly to autoscaling. This is expected since FaaSCache's container eviction policy is only triggered when server resources are exhausted, and is not designed for typical cloud deployments, where resources are plentiful, as is the case in the Azure function traces we use~\cite{shahrad2020serverless}. Therefore, FaaSCache falls back to a conservative dynamic auto-scaling policy. The histogram-based method in~\cite{shahrad2020serverless} and IceBreaker~\cite{roy2022icebreaker} use the function invocation inter-arrival time distribution to predict future invocations, and dynamically pre-warm and keep-alive containers. They outperform autoscaling and further eliminate 13\%--17\% of cold starts. However, neither the histogram model nor IceBreaker's Fourier-transformation-based model can capture complex timeseries patterns nor do they exploit external features, including time of day/week, to improve accuracy.

\SCHEDULER uses the hybrid Bayesian model to account for both timeseries information and external features, and eliminates 24\% more cold starts than IceBreaker, resulting in a cold start rate of less than 4\%.

\begin{figure}
    \centering 
    \includegraphics[width=0.98\columnwidth]{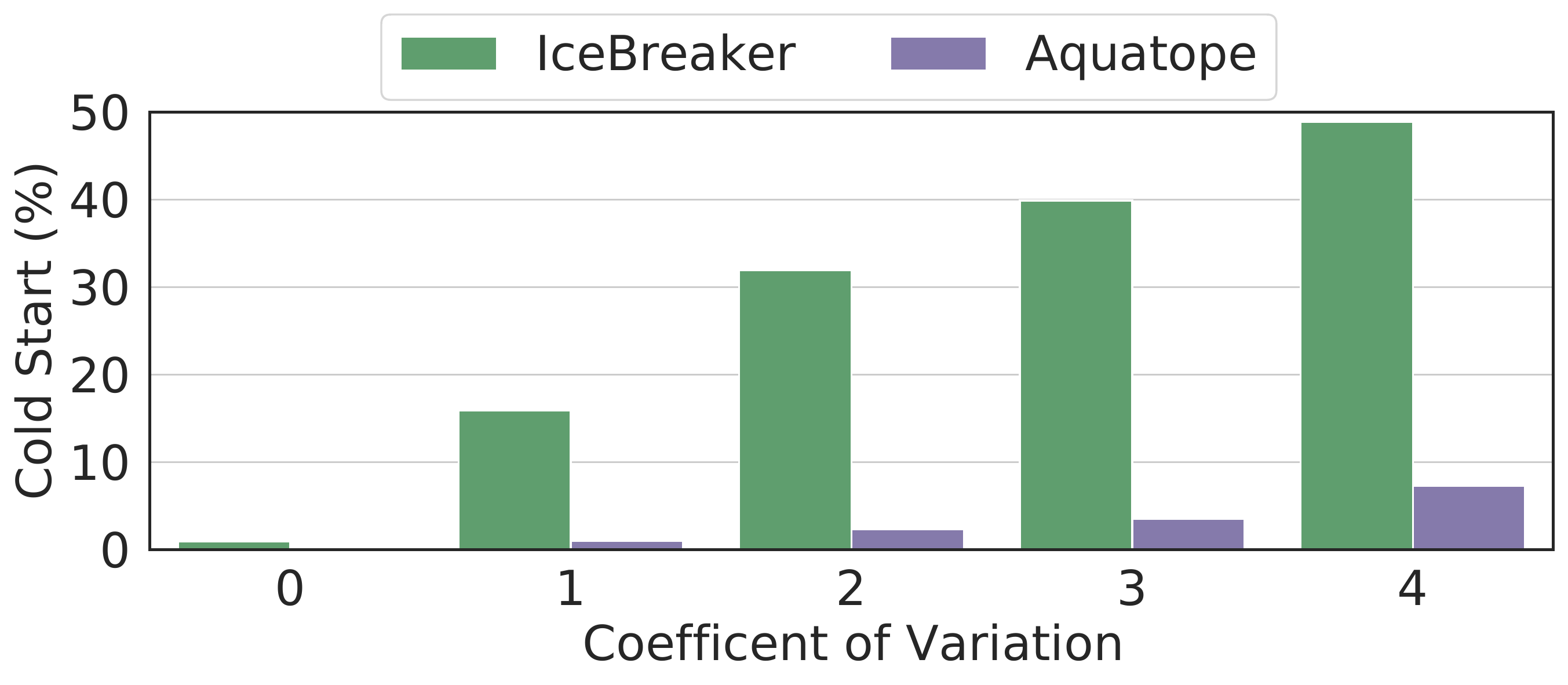}
	\caption{
	Aquatope outperforms IceBreaker, the best-performing previous work for cold start elimination, for input workloads with different coefficients of variation (CV). A CV greater than 1 suggests a large variation in the inter-arrival time of workflow invocations. Aquatope achieves higher benefits for highly fluctuating loads because it accounts for noise and uncertainty.
	}
    \label{fig:eval_cold_cv}
\end{figure}

\begin{figure}
    \centering 
    \includegraphics[width=\columnwidth]{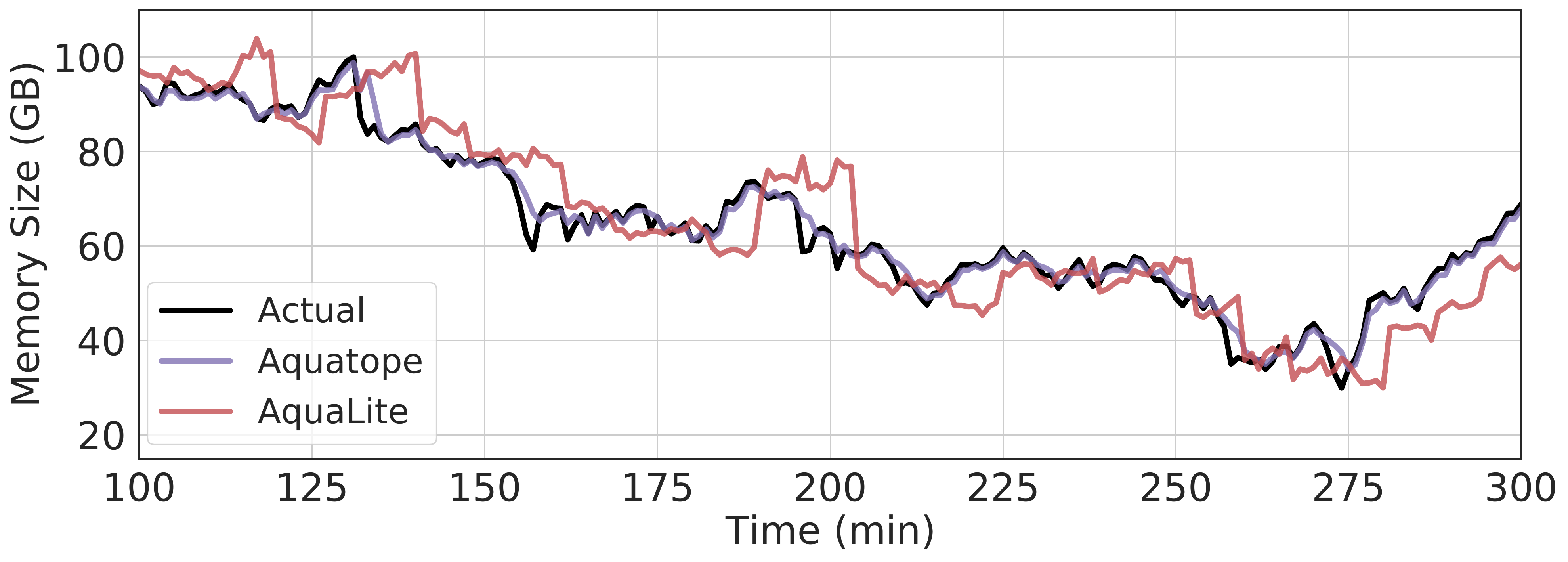}
	\caption{\SCHEDULER's dynamic pre-warmed container pool adapts to fluctuating workload better than the AquaLite that does not account for uncertainty. }
    \label{fig:eval_lstm}
\end{figure}

\begin{figure*}[t]
\centering
    \includegraphics[width=2.05\columnwidth]{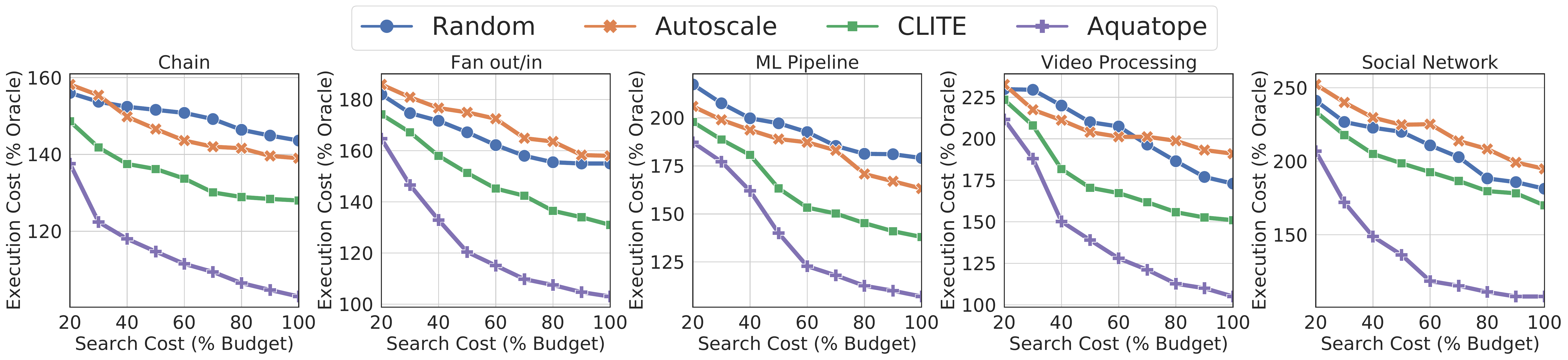}
	\caption{Iteratively searching for a near-optimal configuration across two synthetic and the three end-to-end workflows. }
\label{fig:eval_bo_iter}
\end{figure*}

\vspace{0.03in}
\noindent{\bf{Reducing over-provisioned memory:}}
Although pre-warming containers reduces cold starts, holding containers in memory for too long wastes resources. Fig.~\ref{subfig:eval_wasted_memory} shows the relative aggregate provisioned memory time for each approach. We use the same resource configuration for serverless containers across all approaches for a fair comparison. 

Autoscaling increases the pre-warmed containers in large steps to satisfy performance, but reduces them in much smaller steps when container utilization is low.
However, the temporal bursts common in serverless invocations can lead to over-provisioning of pre-warmed containers, which can take a long time to reclaim resources. As a result, the provisioned memory time of autoscaling is 5\% higher than for \textit{Keep-Alive}. IceBreaker reduces memory time by 25\% compared to \textit{Keep-Alive} by terminating pre-warmed containers right after invocations complete. \SCHEDULER's hybrid prediction model allows it to make fine-grained and timely adjustments to the container pool, and reduces memory time by 23\% compared to IceBreaker.

\vspace{0.03in}
\noindent{\bf{Handling fluctuating load:}}
Aquatope’s dynamic pre-warmed container pool is designed to be noise-aware, making it robust to fluctuating workloads. For the Azure dataset we use, we look at the benefits of Aquatope compared to the best-performing previous work, IceBreaker~\cite{roy2022icebreaker}, for loads with different coefficients of variation (CV) (standard deviation divided by the mean), as shown in Fig.~\ref{fig:eval_cold_cv}. CV greater than 1 indicates significant variability in inter-arrival time~\cite{shahrad2020serverless}. For traces with CVs close to 0, Aquatope yields marginal improvement over IceBreaker. For traces with CV=1 to 4, Aquatope reduces 13\%--41\% more cold starts than IceBreaker, demonstrating the effectiveness of Aquatope's noise-aware approach. In the Azure dataset, more than 40\% of invocation traces have CVs greater than 2, which highlights the high variability present in FaaS environments. 

To further demonstrate the benefits of incorporating noise and uncertainty into the Bayesian prediction model, we also compare \SCHEDULER with a simplified implementation without the uncertainty estimation of Sec.~\ref{sec:hybrid_model}, referred to as AquaLite. The results are shown in Fig.~\ref{fig:eval_lstm}, which shows the aggregate container memory provisioned by AquaLite and \SCHEDULER over time, under a fluctuating load. Thanks to the uncertainty estimation, \SCHEDULER is robust to fluctuating workloads and adjusts the pre-warmed container pool more accurately than AquaLite, reducing 3\% more cold starts and saving 8\% more provisioned memory. 

\vspace{0.03in}
\noindent{\bf{Overhead:}}
\SCHEDULER's container pool scheduler makes adjustment to pre-warmed containers asynchronously, off the critical path, and does not impact the latency of function invocations. Training the hybrid model with a week's trace from Azure Function Dataset~\cite{shahrad2020serverless} takes 50s, which can easily accommodate retraining if needed. The latency of the prediction is below 10ms, which is marginal compared to the adjustment interval of the container pool.


\subsection{Container Resource Manager}
\label{subsec:eval_bo}

\noindent{\bf{Resource efficiency of \SCHEDULER:}}
We first evaluate the resource efficiency of \SCHEDULER's container resource manager, by comparing it with other resource mangers, including Random~\cite{herodotou2011starfish}, Autoscaling~\cite{Autoscale,AutoscaleLimit} and CLITE~\cite{patel2020clite}, in which Random is the baseline policy that randomly selects sample configurations, Autoscaling is a widely adopted resource manager than adjusts resource allocation based on usage, and CLITE is the state-of-the-art BO-driven cloud resource manager that uses a manually crafted objective function to capture the goal of meeting QoS for latency-critical jobs, while maximizing performance for background jobs. We adopt CLITE to the FaaS setting by rewriting its objective function to minimize cost while satisfying QoS.
In our experiments, a QoS violation is defined as failing to meet the end-to-end latency requirement of a serverless workflow. The QoS constraint is chosen to be the latency before saturation is reached, consistent with previous work~\cite{chen2019parties,Zhang21}. We have also conducted experiments with more or less conservative QoS settings and arrived at similar conclusions.

\begin{figure}[t]
\centering
      \includegraphics[width=1\columnwidth]{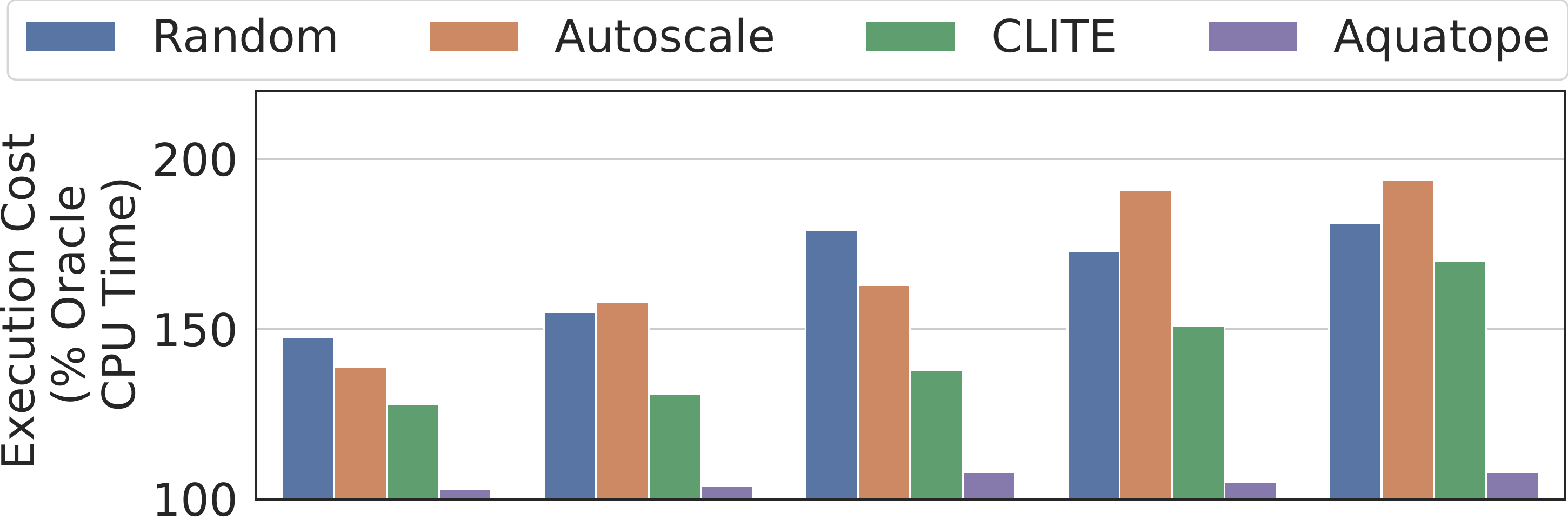}
      \includegraphics[width=1\columnwidth]{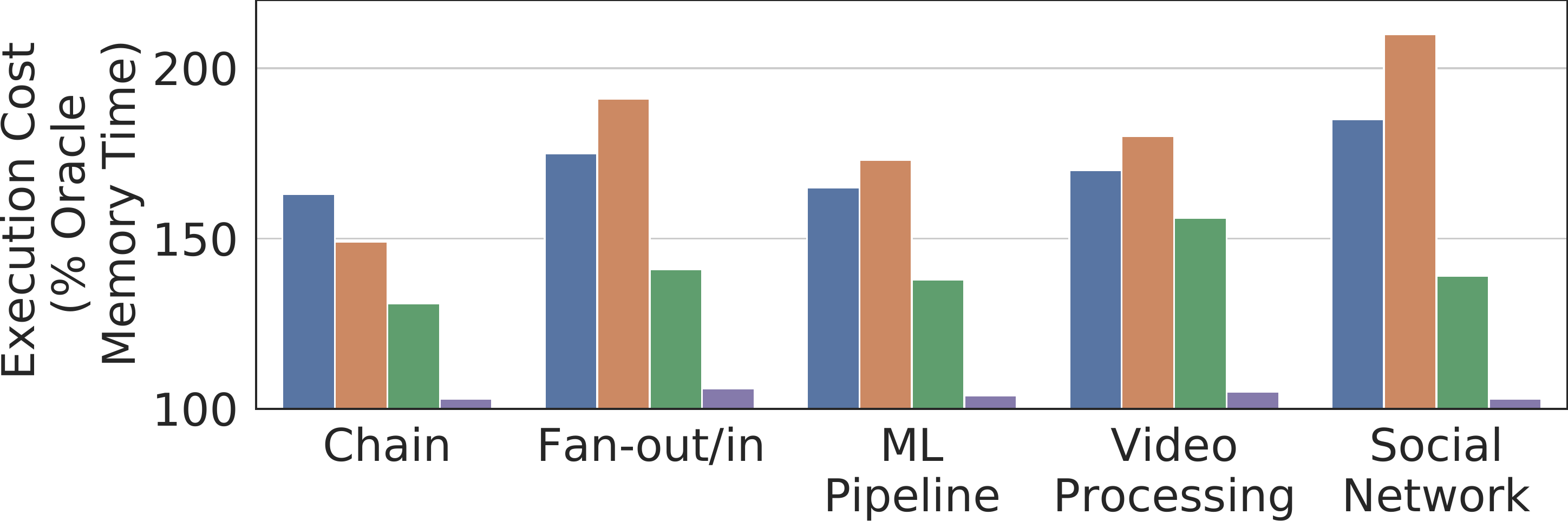}
\caption{\SCHEDULER's resource manager finds a near-optimal resource configuration across multi-stage serverless applications. }
\label{fig:eval_bo_cpu_memory}
\end{figure}

Fig.~\ref{fig:eval_bo_cpu_memory} shows the mean aggregated CPU and memory time of different serverless workflows, under different resource managers. Experiments are repeated 30 times, to account for system noise. For random search, we take the best of all 30 trials for evaluation, because each trial does not always find a QoS-satisfying configuration, consistent with how random search is used in prior work~\cite{alipourfard2017cherrypick}, and all the other resource managers successfully meet QoS. Under the same time budget for resource exploration, \SCHEDULER outperforms all other approaches across examined applications, and significantly reduces CPU and memory time. On average, \SCHEDULER finds a near-optimal configuration with cost within 5\% of the optimal configuration obtained by ORACLE, which exhaustively searches the entire allocation space. As shown in Fig.~\ref{fig:eval_bo_cpu_memory}, \SCHEDULER is not only capable of managing resources for simple applications (e.g., Chain), but can also find near-optimal configurations for complex applications (e.g., Social Network), whose functions vary widely in resource needs. \SCHEDULER outperforms the second best resource managers, using 25\%--62\% less CPUs and 18\%--51\% less memory.

Specifically, Random selects a number of configurations to explore randomly for all stages and never learns from previous trials. In contrast, \SCHEDULER uses a Gaussian process to model the performance of an application based on sampled configurations, and uses prior knowledge to explore the space.
Autoscaling leads to increased cost for two reasons. First, it does not take into account the correlation between execution time and cost of serverless workflow. 
Adding resources can accelerate the computation but also raises the cost per unit of execution time.
Second, it adds resources to all containers belonging to a serverless workflow, rather than only to those that need more resources, leading to overprovisioning. CLITE also results in sub-optimal cost because its manually crafted objective function does not capture the behavior of complex serverless workflows, and often gets trapped in local optima.

\begin{figure}[t]
\centering
\begin{minipage}{.48\linewidth}
  \centering
  \includegraphics[width=1.0\linewidth]{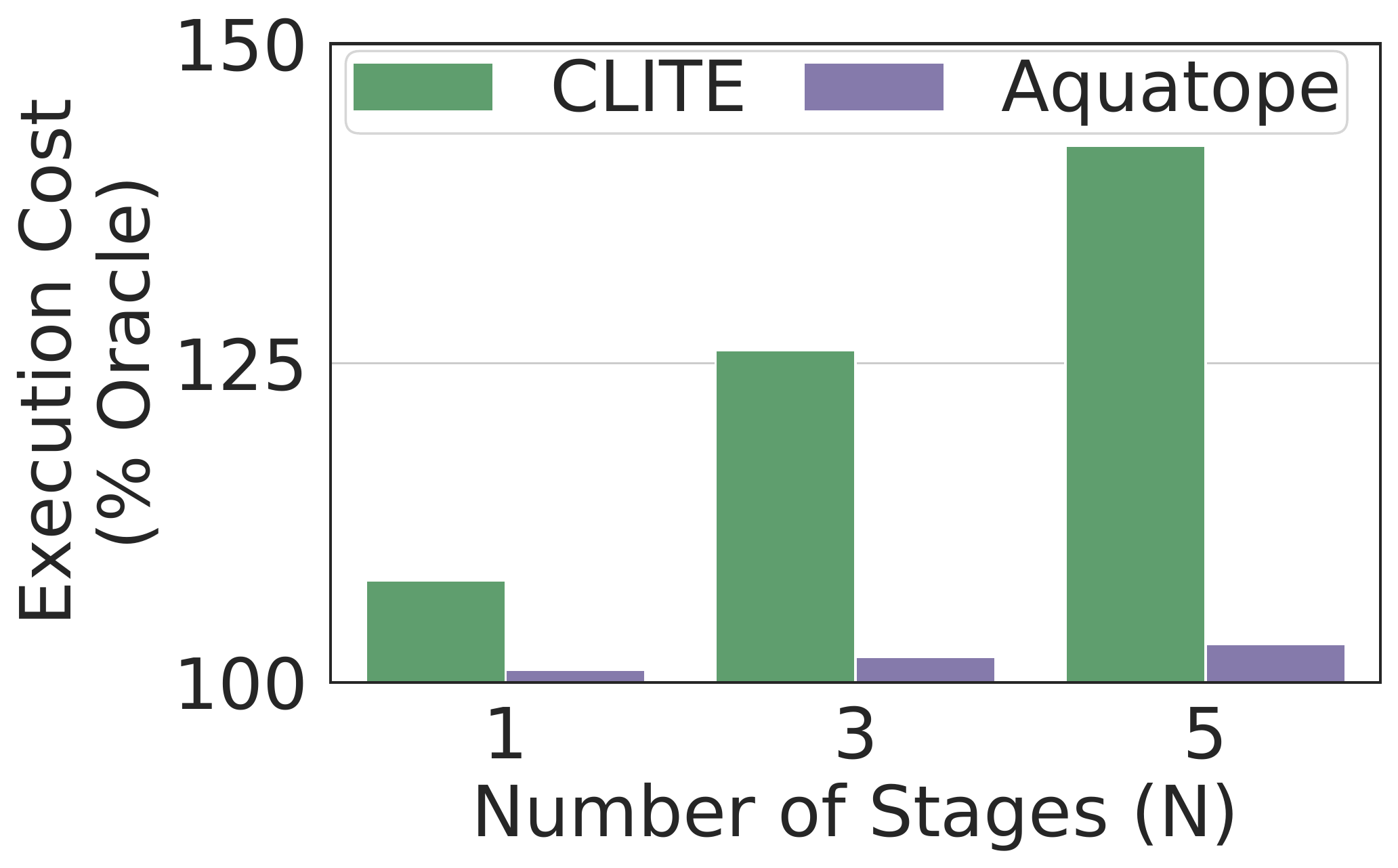}\\
  \subcaption{Function chain.}
  \label{subfig:eval_bo_stage}
\end{minipage}
\begin{minipage}{.48\linewidth}
  \centering
  \includegraphics[width=1.0\linewidth]{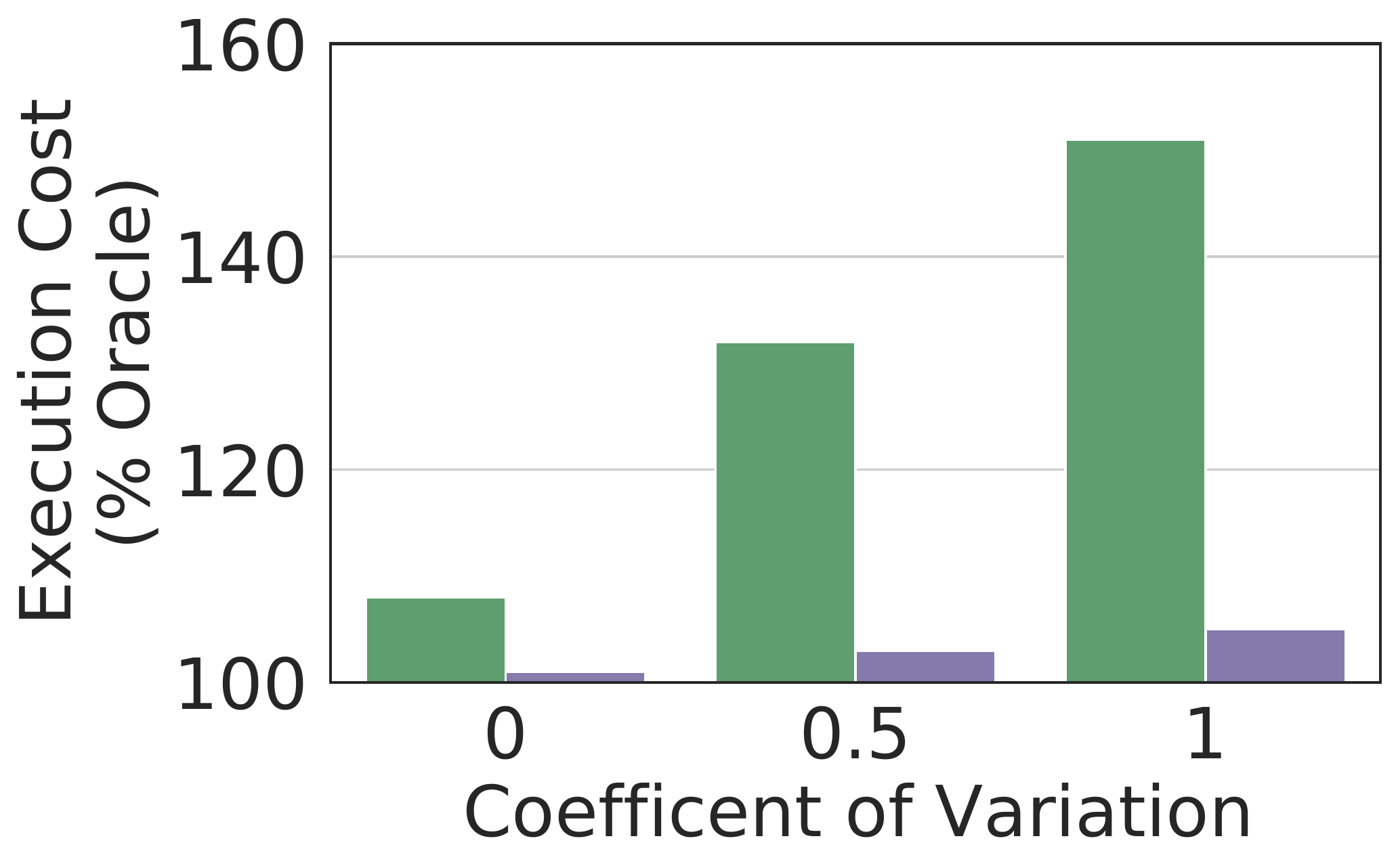}\\
  \subcaption{Varied execution time. }
  \label{subfig:eval_bo_cv}
\end{minipage}
\caption{
Aquatope's resource manager outperforms CLITE~\cite{patel2020clite}, the previous best-performing BO-driven approach, for (a) a function chain with varied number of stages, and (b) a single function workflow with varying degrees of execution time variability.
}
\label{fig:eval_bo_stage_cv}
\end{figure}


\vspace{0.03in}
\noindent{\bf{Fast and accurate convergence:}}
With the customized surrogate models and acquisition functions, \SCHEDULER is able to converge faster and more accurately than other BO-based resource managers, like CLITE, by proactively identifying configurations that may violate QoS and avoiding sampling them.
In addition, \SCHEDULER's batch exploration also yields a substantial reduction in exploration time. As a result, compared to CLITE, \SCHEDULER only spends 31\% wall-clock time on average, and can find a configuration with 36\% lower cost. 
\SCHEDULER also converges more accurately, yielding better resource configurations. Fig.~\ref{fig:eval_bo_iter} shows the resulting cost of all evaluated resource managers for all serverless workflows at different budget levels, and \SCHEDULER constantly converges to the most efficient resource configurations.

\vspace{0.03in}
\noindent{\bf{End-to-end QoS constraint:}}
Aquatope handles end-to-end QoS constraints for complex workflows better than CLITE, which is the best-performing and most closely related previous work. CLITE is designed for colocated monolithic applications or multi-tier applications with defined per-tier QoS targets. However, defining per-tier QoS is a major challenge in real deployments, and most production services do not have per-tier targets. They instead define QoS only based on end-to-end latency. CLITE's hand-crafted  objective function cannot capture the end-to-end performance behavior of complex workflows consisting of multiple functions, whereas Aquatope's independent performance model treats the workflow as a whole, and converges faster and more accurately. As shown in Fig.~\ref{subfig:eval_bo_stage}, when increasing the number of chained functions in a synthetic workflow, Aquatope outperforms CLITE in terms of execution cost by 7\%--39\%. This indicates that Aquatope is better at handling serverless workflows with complex topologies and end-to-end QoS constraints.

\begin{figure}[t]
    \centering
    \includegraphics[width=0.98\columnwidth]{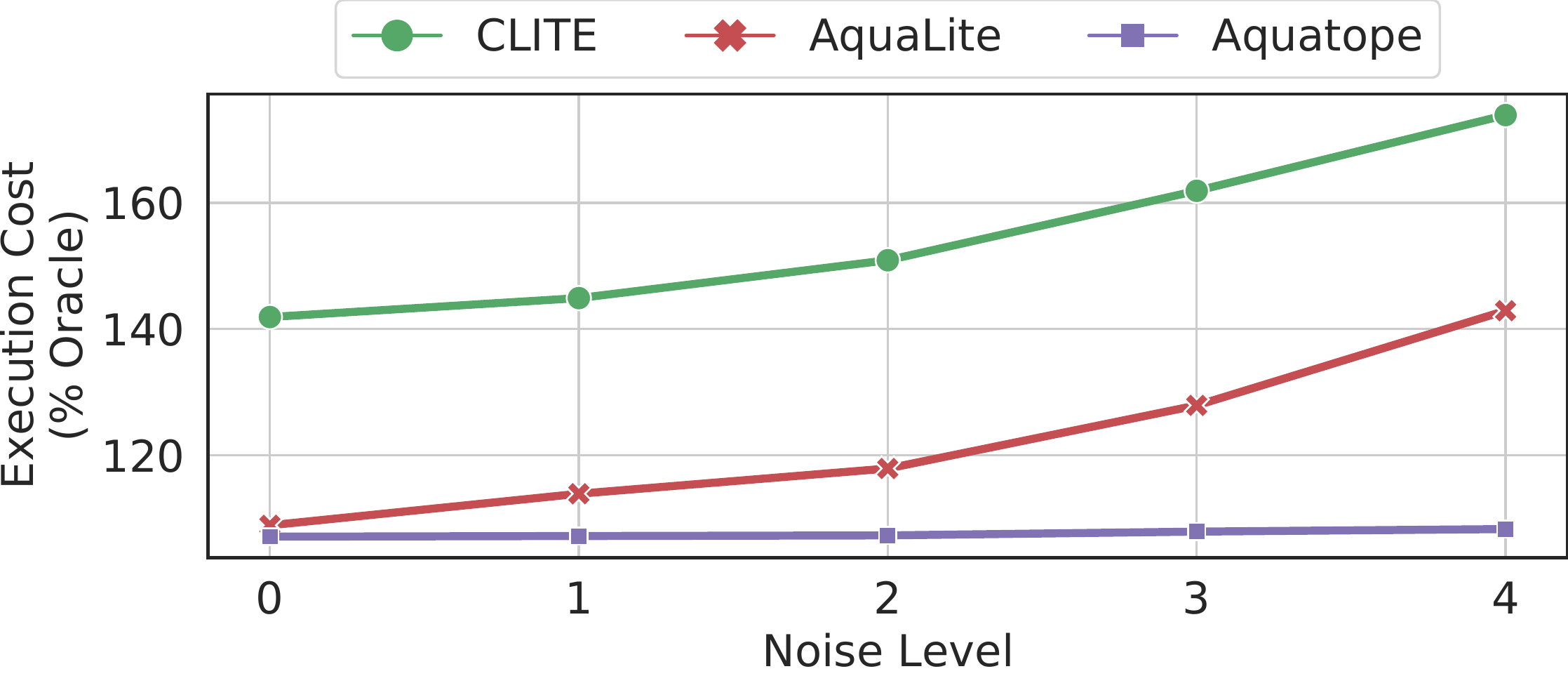}
	\caption{\SCHEDULER's robustness to cloud noise.}
    \label{fig:eval_anomaly}
\end{figure}

\begin{figure}[t]
    \centering 
    \includegraphics[width=0.93\columnwidth]{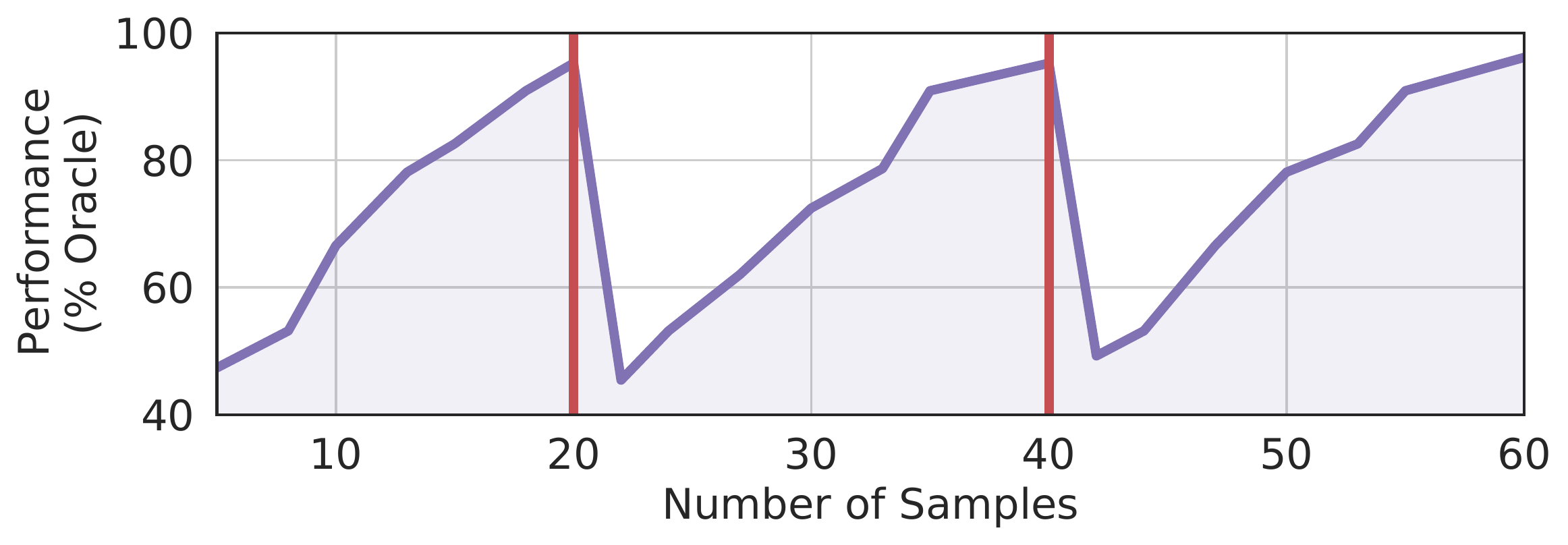}
	\caption{\SCHEDULER adapts to changes in the performance model of the serverless workflow.}
    \label{fig:eval_bo_dynamic}
\end{figure}

\vspace{0.03in}
\noindent{\bf{Resilience to cloud noise:}} 
A major challenge when applying Bayesian Optimization to FaaS is the noise in cloud environments, due to e.g., resource contention. If noise is not handled appropriately, resource managers can violate QoS and/or waste resources. While baseline BO can account for some noise, that is not sufficient to capture the variability of FaaS infrastructures.
Aquatope's resource manager uses customized noise-aware BO to find near-optimal resource configurations under noisy observations. We use a synthetic single function workflow with different degrees of execution time variability to evaluate the performance for Aquatope in a noisy environment. Fig.~\ref{subfig:eval_bo_cv} shows that Aquatope outperforms CLITE in execution cost by 7\%--45\% as the inherent noise of the function increases.

As shown in Fig.~\ref{fig:eval_anomaly}, we further evaluate the robustness of \SCHEDULER to irregular system noise by introducing intermittent background jobs~\cite{tapti2016cloudsuite, ferdman2012cloudsuite} on the same worker servers, causing noise and data outliers in the sampling process of the ML pipeline. The noise level represents the frequency and intensity of the background jobs. As the noise level increases, the number of data outliers increases and the resource manager is more likely to suffer from biased observations. Fig.~\ref{fig:eval_anomaly} illustrates that \SCHEDULER is still able to achieve a near-optimal configuration in the presence of noise and outliers, while CLITE experiences 37--64\% increase in cost. We also compare \SCHEDULER with AquaLite, a simplified version of \SCHEDULER  without the noise-aware components, and find that AquaLite experiences a 10--33\% higher cost compared to \SCHEDULER. This demonstrates the effectiveness of Aquatope by incorporating uncertainty into the performance model and proactively pruning data outliers.

\vspace{0.03in}
\noindent{\bf{Automatic retraining:}}
\SCHEDULER can detect and adapt to changes in performance behavior, which can be caused, for example, by changes in the function inputs or function updates. As shown in Fig.~\ref{fig:eval_bo_dynamic}, \SCHEDULER detects the change in performance behavior when the format and size of the inputs for the video processing pipeline change (marked by red lines), and updates the model dynamically by collecting new samples using a sliding window approach. \SCHEDULER adapts to changes quickly with around 20 new samples within 2 minutes, and is always able to find a new near-optimal resource configuration. 

\vspace{0.03in}
\noindent{\bf{Overhead:}}
\SCHEDULER's container resource manager is not in the critical path of function invocations. Functions continue to execute using their previous resource allocation configuration until \SCHEDULER updates them. The computational overhead of \SCHEDULER is negligible. The time to find the next batch of candidate configurations is less than 100ms, which can be masked by the time needed to evaluate the current samples. 


\subsection{End-to-End Performance}

\begin{figure}[t]
    \centering
    \includegraphics[width=0.74\columnwidth]{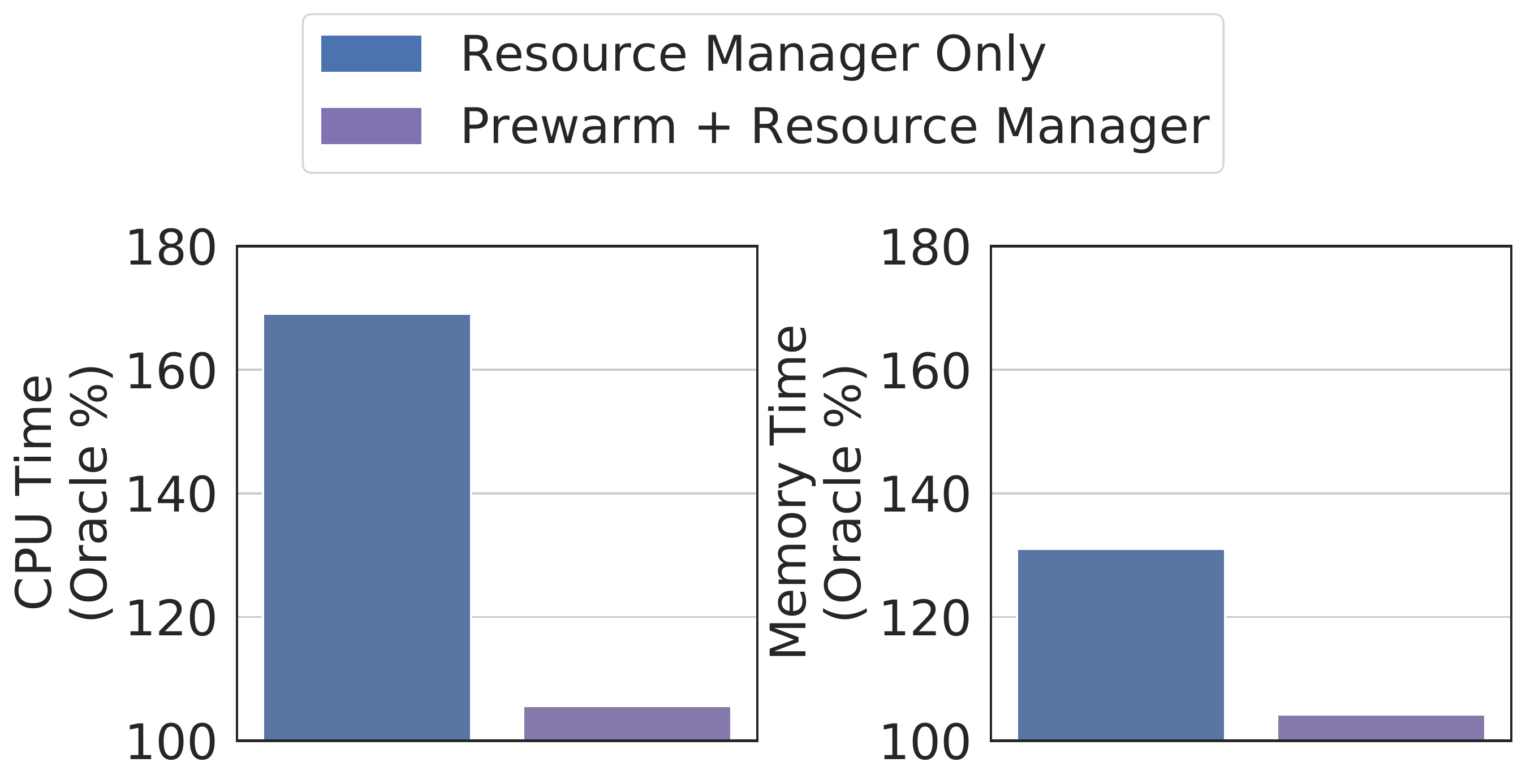}
	\caption{The performance impact of not having the pre-warmed container pool. }
    \label{fig:eval_correlation}
\end{figure}

We first demonstrate that cold starts and resource usage are correlated, and therefore, cold start elimination and resource management need to be tackled jointly. Then we perform an end-to-end evaluation of \SCHEDULER, including both the pre-warmed container pool and resource manager.

We demonstrate the aforementioned correlation, by showing that the resource manager cannot achieve the desired performance without reducing the resource allocation search space to correspond only to warm start containers.
Fig.~\ref{fig:eval_correlation} shows the resulting average CPU and memory time of a fully fledged \SCHEDULER with both the pre-warmed container pool and resource manager, and a simplified \SCHEDULER with only the resource manager in place, compared to the offline oracle. Compared to the fully fledged \SCHEDULER, the simplified version experiences a 64\% increase in CPU time and 28\% increase in memory time. This is due to the diverse behavior of cold and warm starts leading to different resource requirements, and the simplified version of \SCHEDULER being forced to strike a balance between them, leading to degraded performance. This indicates the necessity of jointly tackling cold starts and resource management in FaaS. 

\begin{figure}[t]
    \centering
    \includegraphics[width=1.00\columnwidth]{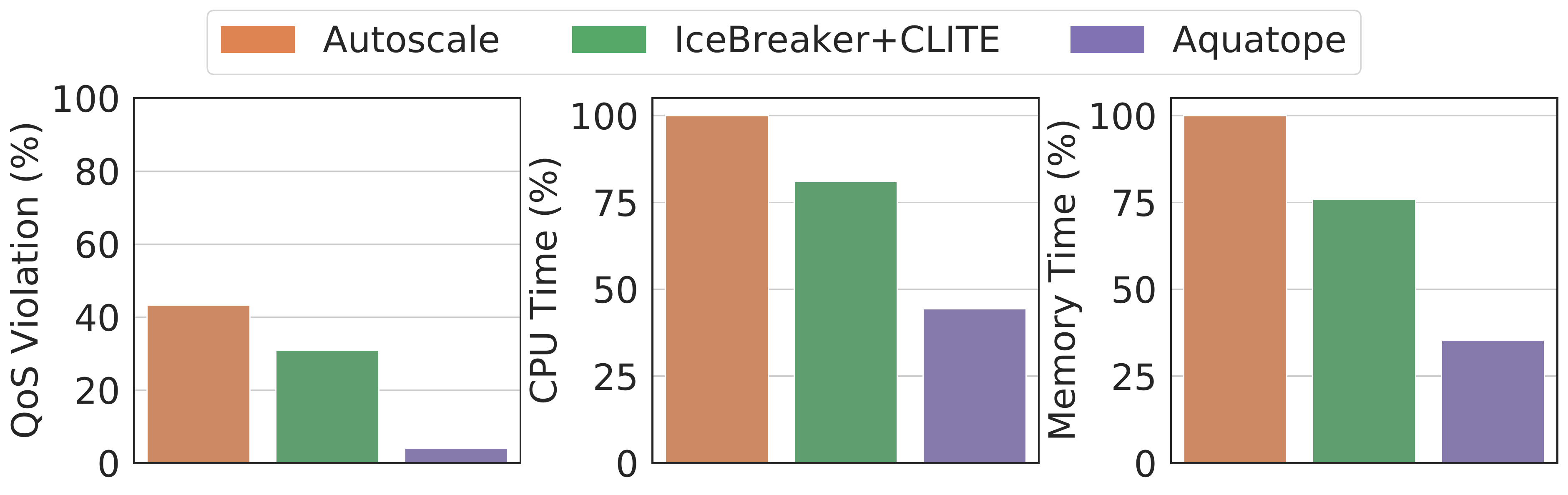}
	\caption{End-to-end performance analysis for \SCHEDULER compared to autoscaling policies and a combination of the best prior work. }
    \label{fig:eval_end_to_end}
\end{figure}


We then perform an end-to-end analysis of the full-fledged \SCHEDULER. Specifically, we compare \SCHEDULER to a framework using the autoscaling-based FaaS resource manager~\cite{openwhiskreactive, lambdascaling, suresh2020ensure} that scales both pre-warmed containers and allocated resources, and a framework combining the container pre-warming mechanism in IceBreaker~\cite{shahrad2020serverless} with the BO-based resource manager in CLITE~\cite{patel2020clite} (IceBreaker+CLITE), which are the best-performing alternatives based on Section~\ref{subsec:prewarm}-\ref{subsec:eval_bo}.
In our experiments, the average CPU utilization is 43\% and the average memory utilization is 29\%, which is consistent with the resource utilization of production clusters~\cite{shan2018lego, zhang2021harvested}. Fig.~\ref{fig:eval_end_to_end} shows the total CPU and memory time of all evaluated frameworks. IceBreaker+CLITE outperforms autoscaling by reducing 13\% of QoS violations, 19\% of CPU time, and 25\% of memory time. In contrast, \SCHEDULER:
\begin{enumerate}
\item Outperforms other approaches, eliminating another 27\%--39\% of the QoS violations, and bringing the total to below 3\%. 
\item Significantly reduces CPU and memory usage, reducing CPU time by 37\%--55\%, and memory time by 41\%--64\%. 
\end{enumerate}
\SCHEDULER achieves these benefits by jointly tackling cold start elimination and resource management, and using Bayesian models that adjust to the behavior of a given application, while remaining general and robust to cloud noise.

%% file: Conclusions.tex
\section{Conclusion}
\label{sec:conclusion}

We have presented \SCHEDULER, a QoS-and-uncertainty-aware resource manager for multi-stage serverless workflows. \SCHEDULER jointly tackles the challenges of cold starts and resource management; the former through the use of a hybrid Bayesian neural network and the latter using customized Bayesian Optimization. Across a diverse set of real-world serverless applications, \SCHEDULER meets QoS, while significantly reducing the amount of required resources.

%% file: main.bbl
\begin{thebibliography}{10}

\bibitem{couchdb}
{Apache CouchDB}.
\newblock \url{https://couchdb.apache.org}.

\bibitem{kafka}
{Apache Kafka}.
\newblock \url{https://kafka.apache.org}.

\bibitem{openwhisk}
{Apache OpenWhisk}.
\newblock \url{https://openwhisk.apache.org}.

\bibitem{composer}
{Apache OpenWhisk Composer}.
\newblock
  \url{https://cloud.ibm.com/docs/openwhisk?topic=openwhisk-pkg_composer}.

\bibitem{awslambda}
{AWS Lambda}.
\newblock \url{https://aws.amazon.com/lambda}.

\bibitem{awsstep}
{AWS Step Functions}.
\newblock \url{https://aws.amazon.com/step-functions}.

\bibitem{azuredurable}
{Azure Durable Functions}.
\newblock
  \url{https://docs.microsoft.com/en-us/azure/azure-functions/durable/durable-functions-overview}.

\bibitem{azurefuntion}
{Azure Functions}.
\newblock \url{https://azure.microsoft.com/en-us/services/functions}.

\bibitem{awspractices}
{Best practices for working with AWS Lambda functions}.
\newblock
  \url{https://docs.aws.amazon.com/lambda/latest/dg/best-practices.html}.

\bibitem{googlefunction}
{Google Cloud Functions}.
\newblock \url{https://cloud.google.com/functions}.

\bibitem{openwhiskreactive}
{How prewarmed containers are provisioned with a reactive configuration}.
\newblock
  \url{https://github.com/apache/openwhisk/blob/master/docs/actions.md}.

\bibitem{ibmfunction}
{IBM Cloud Function}.
\newblock \url{https://cloud.ibm.com/functions}.

\bibitem{lambdascaling}
{Lambda function scaling}.
\newblock
  \url{https://docs.aws.amazon.com/lambda/latest/dg/invocation-scaling.html}.

\bibitem{locust}
{Locust}.
\newblock \url{https://locust.io}.

\bibitem{provisioned}
{Managing concurrency for a Lambda function}.
\newblock
  \url{https://docs.aws.amazon.com/lambda/latest/dg/configuration-concurrency.html}.

\bibitem{minio}
{MinIO}.
\newblock \url{https://min.io}.

\bibitem{nginx}
{Nginx}.
\newblock \url{https://www.nginx.com}.

\bibitem{pytorch}
{PyTorch}.
\newblock \url{https://pytorch.org}.

\bibitem{alipourfard2017cherrypick}
Omid Alipourfard, Hongqiang~Harry Liu, Jianshu Chen, Shivaram Venkataraman,
  Minlan Yu, and Ming Zhang.
\newblock Cherrypick: Adaptively unearthing the best cloud configurations for
  big data analytics.
\newblock In {\em Proceedings of the 14th USENIX Conference on Networked
  Systems Design and Implementation}, NSDI'17, page 469–482, USA, 2017.
  USENIX Association.

\bibitem{ao2019sprocket}
Lixiang Ao, Liz Izhikevich, Geoffrey~M. Voelker, and George Porter.
\newblock Sprocket: A serverless video processing framework.
\newblock In {\em Proceedings of the ACM Symposium on Cloud Computing}, SoCC
  '18, page 263–274, New York, NY, USA, 2018. Association for Computing
  Machinery.

\bibitem{Autoscale}
Autoscale.
\newblock \url{https://cwiki.apache.org/cloudstack/autoscaling.html}.

\bibitem{AutoscaleLimit}
Aws autoscaling.
\newblock \url{http://aws.amazon.com/autoscaling/}.

\bibitem{balandat2020botorch}
Maximilian Balandat, Brian Karrer, Daniel~R. Jiang, Samuel Daulton, Benjamin
  Letham, Andrew~Gordon Wilson, and Eytan Bakshy.
\newblock {BoTorch: A Framework for Efficient Monte-Carlo Bayesian
  Optimization}.
\newblock In {\em Advances in Neural Information Processing Systems 33}, 2020.

\bibitem{brochu2010tutorial}
Eric Brochu, Vlad~M Cora, and Nando De~Freitas.
\newblock A tutorial on bayesian optimization of expensive cost functions, with
  application to active user modeling and hierarchical reinforcement learning.
\newblock {\em arXiv preprint arXiv:1012.2599}, 2010.

\bibitem{chen2019parties}
Shuang Chen, Christina Delimitrou, and Jos\'{e}~F. Mart\'{\i}nez.
\newblock Parties: Qos-aware resource partitioning for multiple interactive
  services.
\newblock In {\em Proceedings of the Twenty-Fourth International Conference on
  Architectural Support for Programming Languages and Operating Systems},
  ASPLOS '19, page 107–120, New York, NY, USA, 2019. Association for
  Computing Machinery.

\bibitem{nilanjan20xanadu}
Nilanjan Daw, Umesh Bellur, and Purushottam Kulkarni.
\newblock Xanadu: Mitigating cascading cold starts in serverless function chain
  deployments.
\newblock In {\em Middleware '20}, pages 356--370. {ACM}, 2020.

\bibitem{Delimitrou13}
Christina Delimitrou and Christos Kozyrakis.
\newblock {Paragon: QoS-Aware Scheduling for Heterogeneous Datacenters}.
\newblock In {\em Proceedings of the Eighteenth International Conference on
  Architectural Support for Programming Languages and Operating Systems
  (ASPLOS)}. Houston, TX, USA, 2013.

\bibitem{Delimitrou13d}
Christina Delimitrou and Christos Kozyrakis.
\newblock {QoS-Aware Scheduling in Heterogeneous Datacenters with Paragon}.
\newblock In {\em ACM Transactions on Computer Systems (TOCS), Vol. 31 Issue
  4}. December 2013.

\bibitem{Delimitrou13c}
Christina Delimitrou and Christos Kozyrakis.
\newblock Quasar: Qos-aware and resource-efficient cluster management.
\newblock In {\em Technical Report}. Stanford, CA, July 2013.

\bibitem{Delimitrou14}
Christina Delimitrou and Christos Kozyrakis.
\newblock {Quasar: Resource-Efficient and QoS-Aware Cluster Management}.
\newblock In {\em Proceedings of the Nineteenth International Conference on
  Architectural Support for Programming Languages and Operating Systems
  (ASPLOS)}. Salt Lake City, UT, USA, 2014.

\bibitem{Delimitrou16}
Christina Delimitrou and Christos Kozyrakis.
\newblock {HCloud: Resource-Efficient Provisioning in Shared Cloud Systems}.
\newblock In {\em Proceedings of the Twenty First International Conference on
  Architectural Support for Programming Languages and Operating Systems
  (ASPLOS)}, April 2016.

\bibitem{Delimitrou15}
Christina Delimitrou, Daniel Sanchez, and Christos Kozyrakis.
\newblock {Tarcil: Reconciling Scheduling Speed and Quality in Large Shared
  Clusters}.
\newblock In {\em Proceedings of the Sixth ACM Symposium on Cloud Computing
  (SOCC)}, August 2015.

\bibitem{du20catalyzer}
Dong Du, Tianyi Yu, Yubin Xia, Binyu Zang, Guanglu Yan, Chenggang Qin, Qixuan
  Wu, and Haibo Chen.
\newblock Catalyzer: Sub-millisecond startup for serverless computing with
  initialization-less booting.
\newblock In {\em Proceedings of the Twenty-Fifth International Conference on
  Architectural Support for Programming Languages and Operating Systems},
  ASPLOS '20, page 467–481, New York, NY, USA, 2020. Association for
  Computing Machinery.

\bibitem{ferdman2012cloudsuite}
Michael Ferdman, Almutaz Adileh, Onur Kocberber, Stavros Volos, Mohammad
  Alisafaee, Djordje Jevdjic, Cansu Kaynak, Adrian~Daniel Popescu, Anastasia
  Ailamaki, and Babak Falsafi.
\newblock Clearing the clouds: A study of emerging scale-out workloads on
  modern hardware.
\newblock {\em Proceedings of the Seventeenth International Conference on
  Architectural Support for Programming Languages and Operating Systems}, 2012.

\bibitem{fuerst2021faascache}
Alexander Fuerst and Prateek Sharma.
\newblock Faascache: Keeping serverless computing alive with greedy-dual
  caching.
\newblock In {\em Proceedings of the 26th ACM International Conference on
  Architectural Support for Programming Languages and Operating Systems},
  ASPLOS 2021, page 386–400, New York, NY, USA, 2021. Association for
  Computing Machinery.

\bibitem{gal2016dropout}
Yarin Gal and Zoubin Ghahramani.
\newblock Dropout as a bayesian approximation: Representing model uncertainty
  in deep learning.
\newblock In {\em Proceedings of the 33rd International Conference on
  International Conference on Machine Learning - Volume 48}, ICML'16, page
  1050–1059. JMLR.org, 2016.

\bibitem{gal2016rnn}
Yarin Gal and Zoubin Ghahramani.
\newblock A theoretically grounded application of dropout in recurrent neural
  networks.
\newblock In {\em Proceedings of the 30th International Conference on Neural
  Information Processing Systems}, NIPS'16, page 1027–1035, 2016.

\bibitem{Gan21}
Yu~Gan, Mingyu Liang, Sundar Dev, David Lo, and Christina Delimitrou.
\newblock {Sage: Practical and Scalable ML-Driven Performance Debugging in
  Microservices}.
\newblock In {\em Proceedings of the Twenty Sixth International Conference on
  Architectural Support for Programming Languages and Operating Systems
  (ASPLOS)}, April 2021.

\bibitem{gan2019asplos}
Yu~Gan, Yanqi Zhang, Dailun Cheng, Ankitha Shetty, Priyal Rathi, Nayan Katarki,
  Ariana Bruno, Justin Hu, Brian Ritchken, Brendon Jackson, Kelvin Hu, Meghna
  Pancholi, Yuan He, Brett Clancy, Chris Colen, Fukang Wen, Catherine Leung,
  Siyuan Wang, Leon Zaruvinsky, Mateo Espinosa, Rick Lin, Zhongling Liu, Jake
  Padilla, and Christina Delimitrou.
\newblock An open-source benchmark suite for microservices and their
  hardware-software implications for cloud \& edge systems.
\newblock In {\em Proceedings of the Twenty-Fourth International Conference on
  Architectural Support for Programming Languages and Operating Systems},
  ASPLOS '19, page 3–18, New York, NY, USA, 2019. Association for Computing
  Machinery.

\bibitem{Delimitrou19}
Yu~Gan, Yanqi Zhang, Kelvin Hu, Yuan He, Meghna Pancholi, Dailun Cheng, and
  Christina Delimitrou.
\newblock {Seer: Leveraging Big Data to Navigate the Complexity of Performance
  Debugging in Cloud Microservices}.
\newblock In {\em Proceedings of the Twenty Fourth International Conference on
  Architectural Support for Programming Languages and Operating Systems
  (ASPLOS)}, April 2019.

\bibitem{gardner2014bayesian}
Jacob~R. Gardner, Matt~J. Kusner, Zhixiang Xu, Kilian~Q. Weinberger, and
  John~P. Cunningham.
\newblock Bayesian optimization with inequality constraints.
\newblock In {\em Proceedings of the 31st International Conference on
  International Conference on Machine Learning - Volume 32}, ICML'14, page
  II–937–II–945. JMLR.org, 2014.

\bibitem{gardner2018gpytorch}
Jacob~R. Gardner, Geoff Pleiss, David Bindel, Kilian~Q. Weinberger, and
  Andrew~Gordon Wilson.
\newblock Gpytorch: Blackbox matrix-matrix gaussian process inference with gpu
  acceleration.
\newblock In {\em Proceedings of the 32nd International Conference on Neural
  Information Processing Systems}, NIPS'18, page 7587–7597, Red Hook, NY,
  USA, 2018. Curran Associates Inc.

\bibitem{jian2020gluoncv}
Jian Guo, He~He, Tong He, Leonard Lausen, Mu~Li, Haibin Lin, Xingjian Shi,
  Chenguang Wang, Junyuan Xie, Sheng Zha, Aston Zhang, Hang Zhang, Zhi Zhang,
  Zhongyue Zhang, Shuai Zheng, and Yi~Zhu.
\newblock Gluoncv and gluonnlp: Deep learning in computer vision and natural
  language processing.
\newblock {\em Journal of Machine Learning Research}, 21(23):1--7, 2020.

\bibitem{herodotou2011starfish}
Herodotos Herodotou, Harold Lim, Gang Luo, Nedyalko Borisov, Liang Dong, Fatma
  Cetin, and Shivnath Babu.
\newblock Starfish: A self-tuning system for big data analytics.
\newblock pages 261--272, 01 2011.

\bibitem{hochreiter1997lstm}
Sepp Hochreiter and J\"{u}rgen Schmidhuber.
\newblock Long short-term memory.
\newblock {\em Neural Comput.}, 9(8):1735–1780, November 1997.

\bibitem{hyndman2008forecast}
Rob Hyndman and Yeasmin Khandakar.
\newblock Automatic time series forecasting: The forecast package for r.
\newblock {\em Journal of Statistical Software}, 26, 07 2008.

\bibitem{klimovic2018pocket}
Ana Klimovic, Yawen Wang, Patrick Stuedi, Animesh Trivedi, Jonas Pfefferle, and
  Christos Kozyrakis.
\newblock Pocket: Elastic ephemeral storage for serverless analytics.
\newblock In {\em Proceedings of the 13th USENIX Conference on Operating
  Systems Design and Implementation}, OSDI'18, page 427–444, USA, 2018.
  USENIX Association.

\bibitem{laptev2017time}
Nikolay Laptev, Jason Yosinski, Li~Erran Li, and Slawek Smyl.
\newblock Time-series extreme event forecasting with neural networks at uber.
\newblock In {\em International conference on machine learning}, volume~34,
  pages 1--5, 2017.

\bibitem{Lazarev21}
Nikita Lazarev, Neil Adit, Shaojie Xiang, Zhiru Zhang, and Christina
  Delimitrou.
\newblock {Dagger: Towards Efficient RPCs in Cloud Microservices with
  Near-Memory Reconfigurable NICs}.
\newblock In {\em Proceedings of the Twenty Sixth International Conference on
  Architectural Support for Programming Languages and Operating Systems
  (ASPLOS)}, April 2021.

\bibitem{letham2019constrained}
Benjamin Letham, Brian Karrer, Guilherme Ottoni, and Eytan Bakshy.
\newblock Constrained bayesian optimization with noisy experiments.
\newblock {\em Bayesian Analysis}, 14(2):495--519, 2019.

\bibitem{li2021rambo}
Qian Li, Bin Li, Pietro Mercati, Ramesh Illikkal, Charlie Tai, Michael
  Kishinevsky, and Christos Kozyrakis.
\newblock Rambo: Resource allocation for microservices using bayesian
  optimization.
\newblock {\em IEEE Computer Architecture Letters}, 20(1):46--49, 2021.

\bibitem{mohan2019agile}
Anup Mohan, Harshad Sane, Kshitij Doshi, Saikrishna Edupuganti, Naren Nayak,
  and Vadim Sukhomlinov.
\newblock Agile cold starts for scalable serverless.
\newblock In {\em Proceedings of the 11th USENIX Conference on Hot Topics in
  Cloud Computing}, HotCloud'19, page~21, USA, 2019. USENIX Association.

\bibitem{tapti2016cloudsuite}
Tapti Palit, Yongming Shen, and Michael Ferdman.
\newblock Demystifying cloud benchmarking.
\newblock In {\em 2016 IEEE International Symposium on Performance Analysis of
  Systems and Software (ISPASS)}, pages 122--132, April 2016.

\bibitem{patel2020clite}
Tirthak Patel and Devesh Tiwari.
\newblock Clite: Efficient and qos-aware co-location of multiple
  latency-critical jobs for warehouse scale computers.
\newblock In {\em 2020 IEEE International Symposium on High Performance
  Computer Architecture (HPCA)}, pages 193--206, 2020.

\bibitem{pu2019shuffling}
Qifan Pu, Shivaram Venkataraman, and Ion Stoica.
\newblock Shuffling, fast and slow: Scalable analytics on serverless
  infrastructure.
\newblock In {\em 16th {USENIX} Symposium on Networked Systems Design and
  Implementation ({NSDI} 19)}, pages 193--206, Boston, MA, February 2019.
  {USENIX} Association.

\bibitem{rasmussen2005gaussian}
Carl~Edward Rasmussen and Christopher K.~I. Williams.
\newblock {\em Gaussian Processes for Machine Learning (Adaptive Computation
  and Machine Learning)}.
\newblock The MIT Press, 2005.

\bibitem{ryan2015network}
Ryan~A. Rossi and Nesreen~K. Ahmed.
\newblock The network data repository with interactive graph analytics and
  visualization.
\newblock In {\em AAAI}, 2015.

\bibitem{roy2021satori}
Rohan~Basu Roy, Tirthak Patel, and Devesh Tiwari.
\newblock Satori: Efficient and fair resource partitioning by sacrificing
  short-term benefits for long-term gains<sup>*</sup>.
\newblock In {\em 2021 ACM/IEEE 48th Annual International Symposium on Computer
  Architecture (ISCA)}, pages 292--305, 2021.

\bibitem{roy2022icebreaker}
Rohan~Basu Roy, Tirthak Patel, and Devesh Tiwari.
\newblock Icebreaker: Warming serverless functions better with heterogeneity.
\newblock In {\em Proceedings of the 27th ACM International Conference on
  Architectural Support for Programming Languages and Operating Systems},
  ASPLOS 2022, page 753–767, New York, NY, USA, 2022. Association for
  Computing Machinery.

\bibitem{saha2018emars}
Aakanksha Saha and Sonika Jindal.
\newblock Emars: Efficient management and allocation of resources in
  serverless.
\newblock In {\em 2018 IEEE 11th International Conference on Cloud Computing
  (CLOUD)}, pages 827--830, 2018.

\bibitem{shahrad2019arch}
Mohammad Shahrad, Jonathan Balkind, and David Wentzlaff.
\newblock Architectural implications of function-as-a-service computing.
\newblock In {\em Proceedings of the 52nd Annual IEEE/ACM International
  Symposium on Microarchitecture}, MICRO '52, page 1063–1075, New York, NY,
  USA, 2019. Association for Computing Machinery.

\bibitem{shahrad2020serverless}
Mohammad Shahrad, Rodrigo Fonseca, Inigo Goiri, Gohar Chaudhry, Paul Batum,
  Jason Cooke, Eduardo Laureano, Colby Tresness, Mark Russinovich, and Ricardo
  Bianchini.
\newblock Serverless in the wild: Characterizing and optimizing the serverless
  workload at a large cloud provider.
\newblock In {\em 2020 {USENIX} Annual Technical Conference ({USENIX} {ATC}
  20)}, pages 205--218. {USENIX} Association, July 2020.

\bibitem{shahriari2016taking}
Bobak Shahriari, Kevin Swersky, Ziyu Wang, Ryan~P. Adams, and Nando de~Freitas.
\newblock Taking the human out of the loop: A review of bayesian optimization.
\newblock {\em Proceedings of the IEEE}, 104(1):148--175, 2016.

\bibitem{shan2018lego}
Yizhou Shan, Yutong Huang, Yilun Chen, and Yiying Zhang.
\newblock Legoos: A disseminated, distributed os for hardware resource
  disaggregation.
\newblock In {\em Proceedings of the 13th USENIX Conference on Operating
  Systems Design and Implementation}, OSDI'18, page 69–87, USA, 2018. USENIX
  Association.

\bibitem{snoek2012practical}
Jasper Snoek, Hugo Larochelle, and Ryan~P. Adams.
\newblock Practical bayesian optimization of machine learning algorithms.
\newblock In {\em Proceedings of the 25th International Conference on Neural
  Information Processing Systems - Volume 2}, NIPS'12, page 2951–2959, Red
  Hook, NY, USA, 2012. Curran Associates Inc.

\bibitem{suresh2020ensure}
Amoghavarsha Suresh, Gagan Somashekar, Anandh Varadarajan, Veerendra~Ramesh
  Kakarla, Hima Upadhyay, and Anshul Gandhi.
\newblock Ensure: Efficient scheduling and autonomous resource management in
  serverless environments.
\newblock In {\em 2020 IEEE International Conference on Autonomic Computing and
  Self-Organizing Systems (ACSOS)}, pages 1--10, 2020.

\bibitem{ustiugov2021snapshots}
Dmitrii Ustiugov, Plamen Petrov, Marios Kogias, Edouard Bugnion, and Boris
  Grot.
\newblock Benchmarking, analysis, and optimization of serverless function
  snapshots.
\newblock In {\em Proceedings of the 26th ACM International Conference on
  Architectural Support for Programming Languages and Operating Systems},
  ASPLOS 2021, page 559–572, New York, NY, USA, 2021. Association for
  Computing Machinery.

\bibitem{wang18peeking}
Liang Wang, Mengyuan Li, Yinqian Zhang, Thomas Ristenpart, and Michael Swift.
\newblock Peeking behind the curtains of serverless platforms.
\newblock In {\em Proceedings of the 2018 USENIX Conference on Usenix Annual
  Technical Conference}, USENIX ATC '18, page 133–145, USA, 2018. USENIX
  Association.

\bibitem{Mars13a}
Hailong Yang, Alex Breslow, Jason Mars, and Lingjia Tang.
\newblock Bubble-flux: precise online qos management for increased utilization
  in warehouse scale computers.
\newblock In {\em Proceedings of ISCA}. 2013.

\bibitem{yu2020serverless}
Tianyi Yu, Qingyuan Liu, Dong Du, Yubin Xia, Binyu Zang, Ziqian Lu, Pingchao
  Yang, Chenggang Qin, and Haibo Chen.
\newblock Characterizing serverless platforms with serverlessbench.
\newblock In {\em Proceedings of the ACM Symposium on Cloud Computing}, SoCC
  '20. Association for Computing Machinery, 2020.

\bibitem{zhang2000neural}
Guoqiang~Peter Zhang.
\newblock Neural networks for classification: a survey.
\newblock {\em IEEE Transactions on Systems, Man, and Cybernetics, Part C
  (Applications and Reviews)}, 30(4):451--462, 2000.

\bibitem{zhang2021harvested}
Yanqi Zhang, \'{I}\~{n}igo Goiri, Gohar~Irfan Chaudhry, Rodrigo Fonseca, Sameh
  Elnikety, Christina Delimitrou, and Ricardo Bianchini.
\newblock Faster and cheaper serverless computing on harvested resources.
\newblock In {\em Proceedings of the ACM SIGOPS 28th Symposium on Operating
  Systems Principles}, SOSP '21, page 724–739, New York, NY, USA, 2021.
  Association for Computing Machinery.

\bibitem{Zhang21}
Yanqi Zhang, Weizhe Hua, Zhuangzhuang Zhou, Edward Suh, and Christina
  Delimitrou.
\newblock {Sinan: ML-Based and QoS-Aware Resource Management for Cloud
  Microservices}.
\newblock In {\em Proceedings of the Twenty Sixth International Conference on
  Architectural Support for Programming Languages and Operating Systems
  (ASPLOS)}, April 2021.

\bibitem{zhu2017deep}
Lingxue Zhu and Nikolay Laptev.
\newblock Deep and confident prediction for time series at uber.
\newblock In {\em 2017 IEEE International Conference on Data Mining Workshops
  (ICDMW)}, pages 103--110, Los Alamitos, CA, USA, Nov 2017. IEEE Computer
  Society.

\end{thebibliography}
